\documentclass[default,iicol]{sn-jnl}

\jyear{2022}

\theoremstyle{thmstyleone}

\theoremstyle{thmstyletwo}

\theoremstyle{thmstylethree}

\usepackage{overpic}
\usepackage{color}
\usepackage{colortbl}
\usepackage{marvosym}

\newcommand{\figref}[1]{Fig.~\ref{#1}}
\newcommand{\tabref}[1]{Table~\ref{#1}}
\newcommand{\secref}[1]{Sec.~\ref{#1}}
\newcommand{\eqnref}[1]{Equ.(\ref{#1})}

\def\ie{\emph{i.e.}}
\def\eg{\emph{e.g.}}
\def\etc{\emph{etc}}
\def\etal{{\em et al.}}

\definecolor{mygray}{gray}{.9}
\definecolor{myRed}{RGB}{219, 68, 55}
\definecolor{myGreen}{RGB}{15, 157, 88}
\definecolor{myBlue}{RGB}{66, 133, 244}

\newcommand{\tabincell}[2]{\begin{tabular}{@{}#1@{}}#2\end{tabular}}

\def\ourmodel{PNS+}
\def\ourdataset{SUN-SEG}

\graphicspath{{./Imgs/}{../../figure/}}

\newcommand\blfootnote[1]{
    \begingroup
    \renewcommand\thefootnote{}\footnote{#1}
    \addtocounter{footnote}{-1}
    \endgroup
}

\begin{document}

\title[PNS+]{\textbf{Video Polyp Segmentation: A Deep Learning Perspective}}

\author[1]{\fnm{Ge-Peng} \sur{Ji}$\textsuperscript{$\dagger$}$} 

\author[2]{\fnm{Guobao} \sur{Xiao}$\textsuperscript{$\dagger$}$}

\author[3]{\fnm{Yu-Cheng} \sur{Chou}$\textsuperscript{$\dagger$}$} 

\author[4]{\fnm{Deng-Ping} \sur{Fan}$\textsuperscript{\Letter}$}

\author[5]{\fnm{Kai} \sur{Zhao}}

\author[6]{\fnm{Geng} \sur{Chen}}

\author[4]{\fnm{Luc Van} \sur{Gool}}

\affil[1]{\orgdiv{Research School of Engineering}, \orgname{Australian National University}, \orgaddress{\city{Canberra}, \country{Australia}}}

\affil[2]{\orgdiv{College of Computer and Control Engineering}, \orgname{Minjiang University}, \orgaddress{\city{Fuzhou}, \country{China}}}

\affil[3]{\orgdiv{Department of Computer Science}, \orgname{Johns Hopkins University}, \orgaddress{\city{Baltimore}, \country{USA}}}

\affil[4]{\orgdiv{Computer Vision Lab}, \orgname{ETH Zürich}, \orgaddress{\city{Zürich}, \country{Switzerland}}}

\affil[5]{\orgdiv{Department of Radiological Sciences}, \orgname{UCLA}, \orgaddress{\city{Los Angeles}, \country{USA}}}

\affil[6]{\orgdiv{School of Computer Science and Engineering}, \orgname{NPU}, \orgaddress{\city{Xi'an}, \country{China}}}

\affil[]{\tt Project Page: \href{https://github.com/GewelsJI/VPS}{https://github.com/GewelsJI/VPS}}

\abstract{
We present the first comprehensive video polyp segmentation (VPS) study in the deep learning era.
Over the years, developments in VPS are not moving forward with ease due to the lack of a large-scale dataset with fine-grained segmentation annotations.
To address this issue, we first introduce a high-quality frame-by-frame annotated VPS dataset, named \ourdataset, which contains 158,690 colonoscopy video frames from the well-known SUN-database.
We provide additional annotation covering diverse types,
\ie, \textit{attribute}, \textit{object mask}, \textit{boundary}, \textit{scribble}, and \textit{polygon}.
Second, we design a simple but efficient baseline, named \ourmodel, which consists of a global encoder, a local encoder, and normalized self-attention (NS) blocks.
The global and local encoders receive an anchor frame and multiple successive frames to extract long-term and short-term spatial-temporal representations, which are then progressively refined by two NS blocks.
Extensive experiments show that \ourmodel~achieves the best performance and real-time inference speed (170fps), making it a promising solution for the VPS task.
Third, we extensively evaluate 13 representative polyp/object segmentation models on our \ourdataset~dataset and provide attribute-based comparisons.
Finally, we discuss several open issues and suggest possible research directions for the VPS community.
}


\maketitle

\section{Introduction}
\blfootnote{$^\dagger$ Equal contribution. $\textsuperscript{\Letter}$ Corresponding author.}

As the second most deadly cancer and the third most common malignancy, colorectal cancer (CRC) is estimated to cause millions of incidence cases and deaths yearly.
The survival rate of CRC patients is over $95\%$ at the first stage of the disease but dramatically decreases to lower than $35\%$ at the fourth and fifth stages \cite{bernal2012towards}.
Therefore, early diagnosis of positive CRC cases through screening techniques, such as colonoscopy and sigmoidoscopy, is vital in increasing the survival rate.
For prevention purposes, physicians can remove the colon polyps that risk turning into cancer.
However, this process highly depends on the physicians' experience and suffers from a high polyp missing rate, \ie, $22\%\sim28\%$~\cite{puyal2020endoscopic}.

Recently, artificial intelligence (AI) techniques are applied to the automatic detection of candidate lesion polyps during colonoscopy for physicians.
However, developing AI models with a satisfactory detection rate is still challenging due to two problems: 
\textbf{(a) Limited Annotated Data.} Deep learning models are often hungry for a large-scale video dataset with densely-annotated labels.
Moreover, a community-agreed benchmark is missing for evaluating the approaches' actual performance.
\textbf{(b) Dynamic Complexity.} The colonoscopy usually involves less ideal conditions of camera-moving acquisition, such as the diversity of colon polyps (\eg, boundary contrast, shape, orientation, shooting angle), internal artifacts (\eg, water flow, residue), and imaging degradation (\eg, color distortion, specular reflection).
To this end, we present a systematic study to facilitate the development of deep learning models for video polyp segmentation (VPS).
The main contributions of this work are summarized as follows:
\begin{itemize}
    \item \textbf{VPS Dataset.} We elaborately introduce a large-scale VPS dataset termed \ourdataset, containing $158,690$ frames selected from the SUN-database~\cite{misawa2020development}.
    We provide a variety of labels, including attribute, object mask, boundary, scribble, and polygon.
    These labels can further support the development of colonoscopy diagnosis, localization, and derivative tasks.
    
    \item \textbf{VPS Baseline.} We design a simple but efficient VPS baseline, named \ourmodel, which consists of a global encoder, a local encoder, and two normalized self-attention (NS) blocks.
    The global and local encoders extract long-and short-term spatial-temporal representations from the first anchor frame and multiple successive frames, respectively.
    The NS block dynamically updates the receptive field when coupling attentive cues among extracted features.
    Experiments show that \ourmodel~achieves the best performance on challenging \ourdataset~dataset.

    \item \textbf{VPS Benchmark.} To comprehensively understand VPS development, we conduct the first large-scale benchmark by evaluating 13 cutting-edge polyp/object segmentation approaches.
    Based on the benchmarking results (\ie, five image-based and eight video-based), we argue that the VPS task is not well undertaken and leaves plenty of room for further exploration.
\end{itemize}

A preliminary version of this work was presented in~\cite{ji2021pnsnet}.
In this extended work, we introduce three different contributions.
\begin{itemize}
    \item In \secref{sec:dataset}, we introduce a high-quality densely-annotated VPS dataset, \ourdataset, with five extended labels, \ie, attribute, object mask, boundary, scribble, and polygon.
    \item Based on the normalized self-attention block as in~\cite{ji2021pnsnet}, we propose a global-to-local learning paradigm to realize the modeling of both long-term and short-term dependencies.
    This part is detailed in~\secref{sec:glns_pipeline}.
    \item As shown in~\secref{sec:vps_benchmark}, we construct the first large-scale benchmark on the VPS task, which contains $13$ the latest polyp/object segmentation competitors.
    We highlight several potential research directions based on the above benchmark results and progress in the VPS field.
\end{itemize}

\section{Related Works}
This section reviews the recent efforts in computer-aided polyp diagnosis from the following two aspects: colonoscopy-related datasets (\secref{sec:related_works_datasets}) and approaches (\secref{sec:related_works_approaches}).

\subsection{Colonoscopy-Related Datasets}\label{sec:related_works_datasets}
Recently, several datasets have been collected for the examination of human colonoscopy.
As shown in~\tabref{tab:DatasetSummary}, we summarize some key statistics of $20$ popular datasets and our \ourdataset~dataset.
In light of the task definition, we categorize them into three main-stream partitions.

\begin{table*}[t!]
    \centering
    \footnotesize
    \renewcommand{\arraystretch}{1.0}
    \setlength\tabcolsep{10.5pt}
    \caption{Statistics of existing 20 datasets for human colonoscopy.
    \textbf{\#IMG} = number of images.
    \textbf{\#VID} = number of video sequences.
    \textbf{DL} = densely labeling.
    \textbf{CLS} = classification label.
    \textbf{BBX} = bounding box.
    \textbf{PM} = pixel-level mask.}
    \label{tab:DatasetSummary}
    \begin{tabular}{r||crr|ccccc}
    \hline
    \textbf{DATASET}~~~
    & \textbf{YEAR} & \textbf{\#IMG} & \textbf{\#VID} & \textbf{DL} & \textbf{CLS} & \textbf{BBX} & \textbf{PM} & \textbf{Website} \\
    \hline
    CVC-ColonDB~\cite{bernal2012towards} 
    &2012 & 300 & 13
    &&&&\checkmark
    & \href{Link}{http://mv.cvc.uab.es/projects/colon-qa/cvccolondb} \\
    ETIS-Larib~\cite{silva2014toward} 
    & 2014 & 196 &34 
    &&&&\checkmark
    & \href{Link}{https://polyp.grand-challenge.org/EtisLarib/} \\
    CVC-ClinicDB~\cite{bernal2015wm}
    & 2015 & 612 & 31 
    &&&&\checkmark
    & \href{Link}{https://polyp.grand-challenge.org/CVCClinicDB/} \\
    ColonoscopicDS~\cite{mesejo2016computer}
    & 2016 & - & 76 
    &&\checkmark&&
    & \href{Link}{http://www.depeca.uah.es/colonoscopy_dataset/} \\
    ASU-Mayo~\cite{tajbakhsh2015automated} 
    & 2016 & 36,458 & 38
    &\checkmark&&&\checkmark
    & \href{Link}{https://polyp.grand-challenge.org/AsuMayo/} \\
    CVC-ClinicVideoDB~\cite{giana2017} 
    & 2017 & 11,954 & 18 
    &\checkmark&&\checkmark&
    & \href{Link}{https://endovissub2017-giana.grand-challenge.org/Home/} \\
    CVC-EndoSceneStill~\cite{vazquez2017benchmark} 
    & 2017 & 912 & 44 
    &&&&\checkmark
    & \href{Link}{http://www.cvc.uab.es/CVC-Colon/index.php/databases/cvc-endoscenestill/} \\
    KID2~\cite{koulaouzidis2017kid,iakovidis2018detecting} 
    & 2017 &2,371 &47 
    &&\checkmark&&\checkmark
    &\href{Link}{https://mdss.uth.gr/datasets/endoscopy/kid/} \\
    Kvasir~\cite{pogorelov2017kvasir} 
    & 2017 & 8,000 & - 
    &&\checkmark&&
    & \href{Link}{https://datasets.simula.no/kvasir/} \\
    EDD2020~\cite{ali2020endoscopy} 
    &2020 & 386 & -
    & &\checkmark &\checkmark &\checkmark
    & \href{Link}{https://edd2020.grand-challenge.org/} \\
    SUN-database~\cite{misawa2020development} 
    &2020 &158,690 & 113
    &\checkmark&\checkmark&\checkmark&
    & \href{Link}{http://amed8k.sundatabase.org/} \\
    Hyper-Kvasir~\cite{borgli2020hyperkvasir} 
    & 2020 & 110,079 & 374 
    &&\checkmark&\checkmark&\checkmark
    & \href{Link}{https://datasets.simula.no/hyper-kvasir/} \\
    Kvasir-SEG~\cite{jha2020kvasir} 
    & 2020 & 1,000 & -
    &&&&\checkmark
    & \href{Link}{https://datasets.simula.no/kvasir-seg/} \\
    PICCOLO~\cite{sanchez2020piccolo} 
    & 2020 & 3,433 & 40
    &&\checkmark&&\checkmark
    & \href{Link}{https://www.biobancovasco.org/en/Sample-and-data-catalog/Databases/PD178-PICCOLO-EN.html} \\
    Kvasir-Capsule~\cite{smedsrud2021kvasir} 
    & 2021 &4,741,504 & 117 
    &\checkmark&\checkmark&\checkmark&
    & \href{Link}{https://github.com/simula/kvasir-capsule} \\
    CP-CHILD-A~\cite{wang2020improved} 
    & 2021 & 8,000 & -
    &&\checkmark&&
    & \href{Link}{https://figshare.com/articles/dataset/CP-CHILD_zip/12554042} \\
    CP-CHILD-B~\cite{wang2020improved} 
    & 2021 & 1,500 & - 
    &&\checkmark&&
    & \href{Link}{https://figshare.com/articles/dataset/CP-CHILD_zip/12554042} \\
    LDPolypVideo~\cite{ma2021ldpolypvideo} 
    & 2021 & 40,266 & 160
    &\checkmark&\checkmark&\checkmark&
    &\href{Link}{https://github.com/dashishi/LDPolypVideo-Benchmark}\\
    KUMC~\cite{li2021colonoscopy} 
    & 2021 & 37,899 & 155
    &\checkmark&\checkmark&\checkmark&
    &\href{Link}{https://dataverse.harvard.edu/dataset.xhtml?persistentId=doi:10.7910/DVN/FCBUOR}\\
    PolypGen~\cite{ali2021polypgen} 
    & 2021 & 6,282 & 26
    & &\checkmark &\checkmark &\checkmark
    &\href{Link}{https://github.com/sharibox/PolypGen-Benchmark}\\
    \hline
    \rowcolor{mygray}
    \textbf{\ourdataset~(OUR)} & 2022 & 158,690 &1,013
    &\checkmark &\checkmark &\checkmark &\checkmark
    &\href{Link}{https://github.com/GewelsJI/VPS} \\
    \hline
    \end{tabular}
\end{table*}

\subsubsection{Classification}
There are four popular datasets for the initial purpose of identifying gastrointestinal lesions.
ColonoscopicDS~\cite{mesejo2016computer} collects $76$ regular colonoscopy videos with three types of gastrointestinal lesions, including hyperplasic, serrated, and adenoma lesions.
Kvasir~\cite{pogorelov2017kvasir} contains $8$ types of anatomical landmarks (\ie, polyps, esophagitis, ulcerative colitis, z-line, pylorus, cecum, dyed polyp, and dyed resection margins), and each type has $1,000$ images.
Hyper-Kvasir~\cite{borgli2020hyperkvasir} further collects $110,079$ samples from $374$ colonoscopy videos, containing three types of annotations: $10,662$ class labels with $23$ different lesion findings and $1,000$ images with segmented masks and bounding box labels.
Notably, all the segmented masks in Hyper-Kvasir are selected from Kvasir-SEG~\cite{borgli2020hyperkvasir}.
Recently, CP-CHILD-A \& -B~\cite{wang2020improved} record the colonoscopy data from children, including two classes (\ie, colon polyp, normal or other pathological images) for the classification task.

\subsubsection{Detection}
There are five widely-accepted video datasets mainly used for the detection task.
CVC-ClinicVideoDB~\cite{giana2017}, as the early video dataset, comprises $18$ videos with a total number of $11,954$ frames in which $10,025$ frames contain at least a polyp.
As for the largest densely-annotated video polyp detection dataset, SUN-database~\cite{misawa2020development} consists of $49,136$ positive samples with their bounding boxes acquired from $99$ patients.
More recently, two video datasets (\ie, Kvasir-Capsule~\cite{smedsrud2021kvasir} and KUMC~\cite{li2021colonoscopy}) are applied for both detection and classification tasks.
Especially, the former provides $47,238$ bounding box labels from $14$ lesion classes, and the latter has $37,899$ frames with bounding box labels.
Unlike the above datasets, LDPolypVideo~\cite{ma2021ldpolypvideo} includes $40,266$ frames with circular annotations from $160$ colonoscopy videos.

\subsubsection{Segmentation}
As for the video datasets, the early benchmark CVC-EndoSceneStill~\cite{vazquez2017benchmark} opts for the combination of CVC-ColonDB~\cite{bernal2012towards} and CVC-ClinicDB~\cite{bernal2015wm}.
ETIS-Larib~\cite{silva2014toward} provides $196$ labeled samples from $32$ colonoscopy videos, containing about five frames for each sequence.
EDD2020~\cite{ali2020endoscopy} contains $386$ endoscopy images from five different institutions and multiple gastrointestinal organs.
They provide annotations for disease detection, localization, and segmentation.
PICCOLO~\cite{sanchez2020piccolo} also samples $3,433$ frames from $40$ videos with sparse annotations.
As such, the above five video datasets adopt the sampling annotation strategy, which still lacks per-frame masks on each video sequence due to the labor-intensive annotation process.
Being the pioneering video dataset with densely-annotated masks, ASU-Mayo~\cite{tajbakhsh2015automated} contains $36,458$ continuous frames from $38$ videos, while it only provides $3,856$ labels for $10$ positive videos.
Recently, PolypGen~\cite{ali2021polypgen} has collected a multi-centre dataset incorporating more than $300$ patients, including single and continuous frames with $3,788$ annotated segmentation masks and bounding box labels.
Unlike existing works, we introduce \ourdataset, the first high-quality densely-annotated dataset for the VPS task, which contains rich annotated labels, such as object mask, boundary, scribble, polygon, and attribute.
We hope that this work could fuel the development of colonoscopy diagnosis, localization, and derivative tasks.

\subsection{Colonoscopy-Related Methods}\label{sec:related_works_approaches}
Early solutions~\cite{bernal2012towards,dhandra2006analysis,mamonov2014automated,maghsoudi2017superpixel} have been dedicated to identifying colon polyps via mining hand-crafted patterns, such as color, shape, texture, and super-pixel.
However, they usually suffer from low accuracy due to the limited capability of representing heterogeneous polyps, as well as the close resemblance between polyps and hard mimics~\cite{yu2016integrating}.
In contrast, data-driven AI techniques can handle these challenging conditions with better learning ability.
This section mainly focuses on tracking the latest image/video polyp segmentation techniques~\cite{tavanapong2022artificial}, while leaving the systematic review of polyp classification~\cite{gammulle2020two,carneiro2020deep} and detection~\cite{zhang2018polyp,wu2021multi} in our future works.

\subsubsection{Image Polyp Segmentation (IPS)}
Several methods have been proposed to locate the pixel-level polyp regions from the colonoscopy images.
They can be grouped into two major categories.
\textbf{(a) CNN-based Approaches.}
Brandao \etal~\cite{brandao2017fully} adopted a fully convolutional network (FCN) with a pre-trained model to segment polyps.
Later, Akbari~\etal~\cite{akbari2018polyp} introduced a modified FCN to improve the segmentation accuracy.
Inspired by the vast success of UNet~\cite{ronneberger2015u} in biomedical image segmentation, UNet++~\cite{zhou2018unetplus} and ResUNet~\cite{jha2019resunetplus} were employed for polyp segmentation for improved performance.
Furthermore, PolypSeg~\cite{zhong2020polypseg}, ACS~\cite{zhang2020adaptive}, ColonSegNet~\cite{jha2021real} and SCR-Net~\cite{wu2021precise} explore the effectiveness of UNet-enhanced architecture on adaptively learning semantic contexts.
As the newly-proposed methods, SANet~\cite{wei2021shallow} and MSNet~\cite{zhao2021automatic} design the shallow attention module and subtraction unit, respectively, to achieve precise and efficient segmentation.
Additionally, several works opt for introducing additional constraints via three main-stream manners: exerting explicit boundary supervision~\cite{murugesan2019psi,wang2022boundary,fang2019selective,shen2021hrenet,ji2022eernet}, introducing implicit boundary-aware representation~\cite{fan2020pra,nguyen2021ccbanet,cheng2021learnable}, and exploring uncertainty for the ambiguous regions~\cite{kim2021uacanet}.
\textbf{(b) Transformer-based Approaches.}
Recently, Transformers \cite{shamshad2022transformers} have been gaining popularity thanks to their powerful modeling ability.
TransFuse~\cite{zhang2021transfuse} combines the Transformer and CNN, termed the parallel-in-branch scheme, for capturing global dependencies and low-level spatial details.
Besides, A BiFusion module was designed to fuse multi-level features from both branches.
Segtran~\cite{li2021medical} proposes a squeezed attention block that regularizes self-attention, and the expansion block learns diversified representations.
A positional encoding scheme was proposed to impose an inductive continuity bias.
Based on PVT~\cite{wang2021pvtv2}, Dong~\etal~\cite{dong2021polyp} introduced a model with three tight components, \ie, cascaded fusion, camouflage identification, and similarity aggregation modules.

\subsubsection{Video Polyp Segmentation (VPS)}
Despite their progress, existing IPS methods suffer from an inherent limitation of overlooking the valuable temporal cues in the colonoscopy videos.
Therefore, efforts have been dedicated to combining spatial-temporal features among consecutive video frames.
A hybrid 2/3D CNN framework~\cite{puyal2020endoscopic} was proposed to aggregate spatial-temporal correlations and achieves better segmentation results.
However, the kernel size restricts the spatial correlation between frames, restricting the accurate segmentation of fast movements of polyps.
To alleviate the above problem, PNSNet~\cite{ji2021pnsnet} introduces a normalized self-attention (NS) block to learn spatial-temporal representations with neighborhood correlations effectively.
In this paper, we delve deeper into a more effective global-to-local learning strategy based on NS block, which can fully leverage both long-term and short-term spatial-temporal dependencies.

\section{VPS Dataset}\label{sec:dataset}
We describe the introduced \ourdataset~dataset's details in terms of data collection/re-organization (\secref{sec:data_collection}), professional annotations (\secref{sec:hierarchy_annotation}), and dataset statistics (\secref{sec:dataset_statstics}).

\subsection{Data Organization}\label{sec:data_collection}
The colonoscopy videos in our \ourdataset~are from Showa University and Nagoya University database (also named SUN-database~\cite{misawa2020development}), the largest video polyp dataset for the detection task.
There are two advantages of adopting the SUN-database as our data source.
\textbf{(a) Challenging Scenarios:} 
The videos are captured by the high-definition endoscope (CF-HQ290ZI \& CF-H290ECI, Olympus) and video recorder (IMH-10, Olympus), providing videos of various polyp sizes at dynamic scenarios, such as imaging at different focusing distances and speeds.
\textbf{(b) Reliable Pathological Localization:}
The initial classification information and bounding box annotations are provided by three research assistants and examined by two expert endoscopists with professional domain knowledge.

\begin{figure}[t!]
    \centering
    \includegraphics[width=\linewidth]{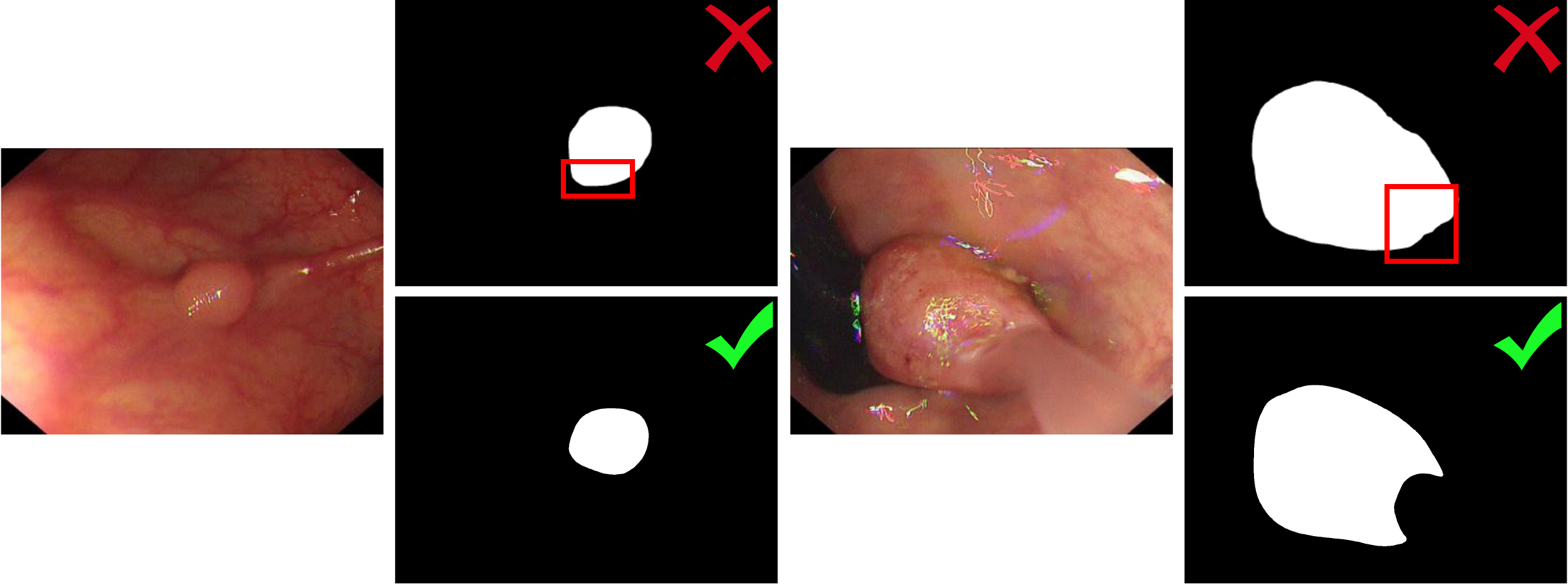}
    \put(-216, 14.5){\footnotesize (a) Redundant}
    \put(-207, 6.5){\footnotesize Annotation}
    \put(-98, 14.5){\footnotesize (b) Water}
    \put(-97, 6.5){\footnotesize Occlusion}
    \caption{High-criteria control for data annotation.
    For instance, we reject case (a), where the boundary is not consistent with the polyp, and case (b), where the water overlapping area is falsely annotated.
    }\label{fig:high_qual_ann}
\end{figure}
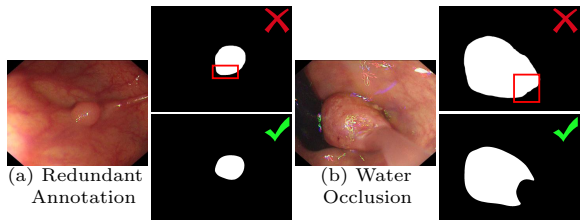

The origin SUN-database has $113$ colonoscopy videos, including $100$ positive cases with $49,136$ polyp frames and $13$ negative cases with $109,554$ non-polyp frames\footnote{These statistic data come from this \href{website}{http://amed8k.sundatabase.org/}, which is different from the data reported in the original paper~\cite{misawa2020development}.
Besides, the SUN-database is available for only non-commercial use in research or educational purpose, which could be freely accessed with permission from authors.}.
We manually trim them into $378$ positive and $728$ negative clips while maintaining their consecutive intrinsic relationship.
Such data pre-processing ensures that each clip has around 3$\sim$11s duration at a real-time frame rate (\ie, $30$ fps), promoting the fault-tolerant margin for various algorithms and devices.
To this end, the re-organized \ourdataset~contains $1,106$ short video clips with $158,690$ video frames totally, offering a solid foundation to build such a representative benchmark.

\subsection{Professional Annotations}\label{sec:hierarchy_annotation}
\begin{figure}[t!]
    \centering
    \includegraphics[width=\linewidth]{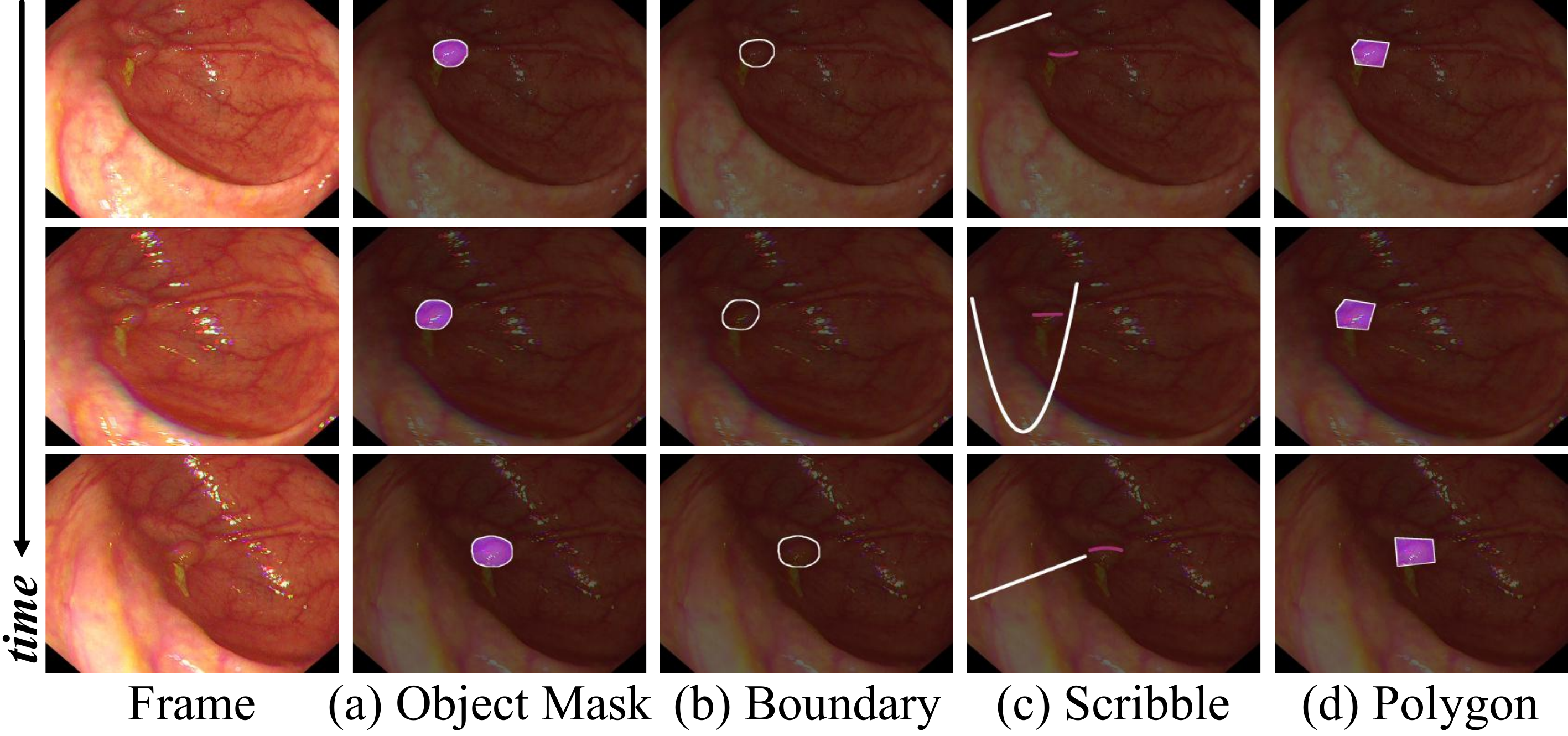}
    \caption{Diversified annotations for each video frame in our \ourdataset~dataset including object mask (a), boundary (b), and two weak labels, \ie, scribble (c) and polygon (d).
    More details refer to~\secref{sec:hierarchy_annotation}.
    }\label{fig:annotation_types}
\end{figure}

Following~\cite{fan2020camouflaged}, we adopt a similar annotation pipeline.
According to the origin bounding box labels of the SUN-database~\cite{misawa2020development}, 
ten experienced annotators are instructed to offer various labels using Adobe Photoshop.
Then, three colonoscopy-related researchers re-verify the quality and correctness of these initial annotations.
\figref{fig:high_qual_ann} shows two typical samples under the restricted quality controls (\ie, rejected and passed).
In addition to the original pathological materials provided by SUN-database, such as pathological pattern (\eg, low-grade adenoma, hyperplastic polyp, \etc), shape (\eg, pedunculated, subpedunculated, \etc), and location (\eg, cecum, ascending colon, \etc), we further extend them with diversified annotations in our \ourdataset.
The newly-extended annotations consist of the following five hierarchies: visual attribute $\rightarrow$ object mask $\rightarrow$ boundary $\rightarrow$ scribble $\rightarrow$ polygon.
Selected samples and corresponding annotations could be found in~\figref{fig:annotation_types} and their illustrations\footnote{The descriptions of complete annotations refer to \url{https://github.com/GewelsJI/VPS/blob/main/docs/DATA_DESCRIPTION.md}.} are as follows.

\begin{table*}[t!]
    \footnotesize
    \renewcommand{\arraystretch}{1.0}
    \setlength\tabcolsep{4pt}
    \caption{List of ten types of visual attributes (ATTR.) and their descriptions.}\label{tab:Attr}
    \begin{tabular*}{\linewidth}{cl}
    \hline
    \textbf{ATTR.} & \textbf{DESCRIPTION} \\
    \hline
    \textbf{SI}
    & \emph{Surgical Instruments.} The endoscopic surgical procedures involve the positioning of instruments, such as snares, \\
    & forceps, knives, and electrodes. \\
    \textbf{IB}
    & \emph{Indefinable Boundaries.} The foreground and background areas around the object have similar color. \\
    \textbf{HO}
    & \emph{Heterogeneous Object.} Object regions have distinct colors. \\
    \textbf{GH}
    & \emph{Ghosting.} Object has anomaly RGB-colored boundary due to fast moving or insufficient refresh rate. \\
    \textbf{FM}
    & \emph{Fast-motion.} The average per-frame object motion in a clip, computed as the Euclidean distance of polyp centroids \\
    & between consecutive frames, is larger than $20$ pixels. \\
    \textbf{SO}
    & \emph{Small Object.} The average ratio between the object size and the image area in a clip is smaller than $0.05$. \\
    \textbf{LO}
    & \emph{Large Object.} The average ratio between the  object size and the image area in a clip is larger than $0.15$. \\
    \textbf{OC}
    & \emph{Occlusion.} Polyp object becomes partially or fully occluded. \\
    \textbf{OV}
    & \emph{Out-of-view.} Polyp object is partially clipped by the image boundaries. \\
    \textbf{SV}
    & \emph{Scale-variation.} The average area ratio among any pair of bounding boxes enclosing the target object in a clip is \\
    & smaller than $0.5$. \\
    \hline
    \end{tabular*}
\end{table*}

\begin{itemize}
    \item \textit{Visual Attribute.} 
    According to the visual characteristics of videos, we provide ten visual attributes at the video level, whose classification criteria are detailed in~\tabref{tab:Attr}.

    \item \textit{Object Mask.}
    Correctly parsing lesion areas is helpful for a clinician.
    Therefore, as shown in~\figref{fig:annotation_types}~(a), we provide pixel-wise object masks for each frame.
    We further refine the coordinates of the original bounding box based on the object mask to tighten the target, offering more reliable localization labels.

    \item \textit{Boundary.}
    \figref{fig:annotation_types}~(b) shows the polyp boundary generated by calculating the gradient of the object mask.

    \item \textit{Scribble.}
    Besides, we offer two weak labels to facilitate the research under data-insufficient conditions.
    As for the scribble labels in \figref{fig:annotation_types}~(c), we use two high-degree curves to indicate the foreground (purple curve) and background (white curve), respectively.
    To ensure the objectivity of various annotators, we adopt linear or quadratic functions to randomly create the above curves in the positive/negative region.

    \item \textit{Polygon.}
    Similarly, in~\figref{fig:annotation_types}~(d), we randomly deploy the Douglas-Peucker algorithm~\cite{RAMER1972244} to find the circumscribed or inscribed polygons that fit the object boundaries.
\end{itemize}

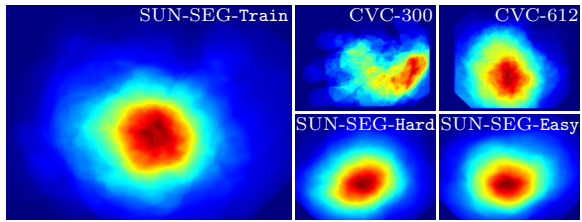
\begin{figure}[t!]
    \centering
    \includegraphics[width=\linewidth]{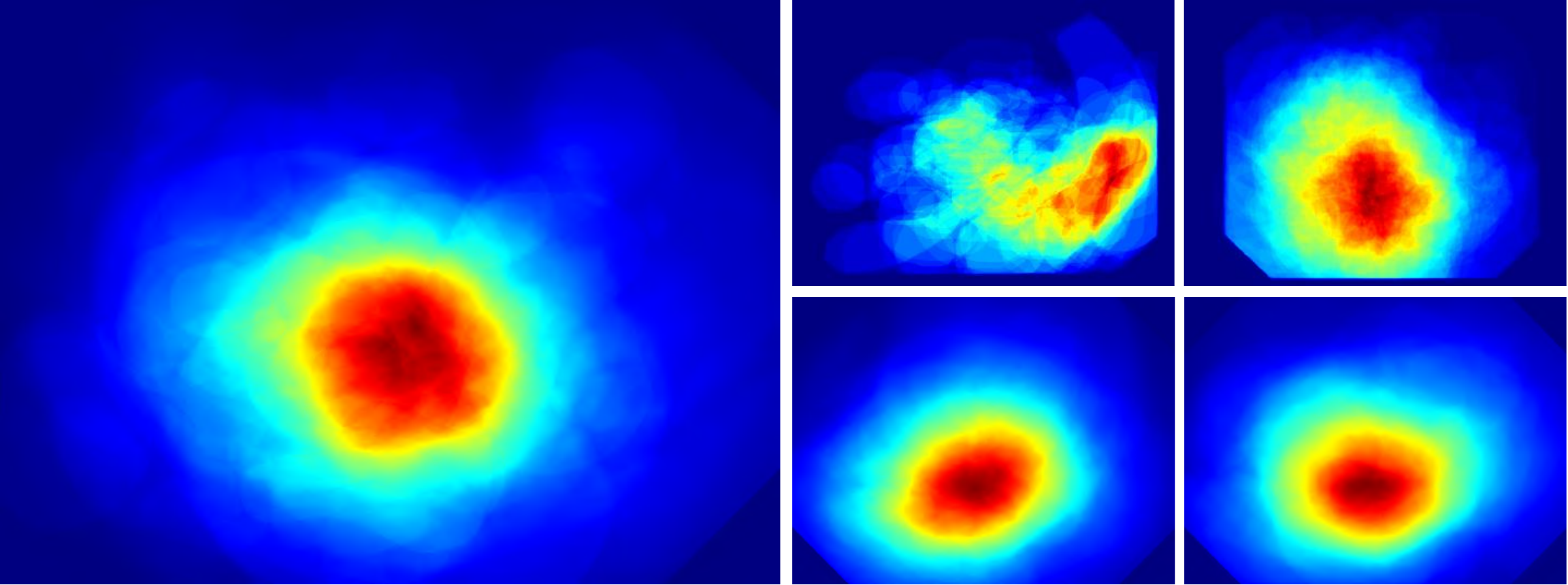}
    \put(-166, 75){\footnotesize \color{white} \ourdataset-\texttt{Train}}
    \put(-87, 75){\footnotesize \color{white} CVC-300}
    \put(-32, 75){\footnotesize \color{white} CVC-612}
    \put(-107, 34){\footnotesize \color{white} \ourdataset-\texttt{Hard}}
    \put(-52.5, 34){\footnotesize \color{white} \ourdataset-\texttt{Easy}}
    \caption{The calculation of center bias~\cite{fan2021salient} on CVC-300, CVC-612, and our \ourdataset-\texttt{Train}/-\texttt{Easy}/-\texttt{Hard}.}
    \label{fig:center_bias}
\end{figure}

\subsection{Dataset Statistics}\label{sec:dataset_statstics}
This section discusses several vital statistics of our three \ourdataset~sub-datasets for better illustration.
More details about the data split of \ourdataset~refer to~\secref{sec:data_split}.

\begin{figure}[!t]
    \centering
    \includegraphics[width=\linewidth]{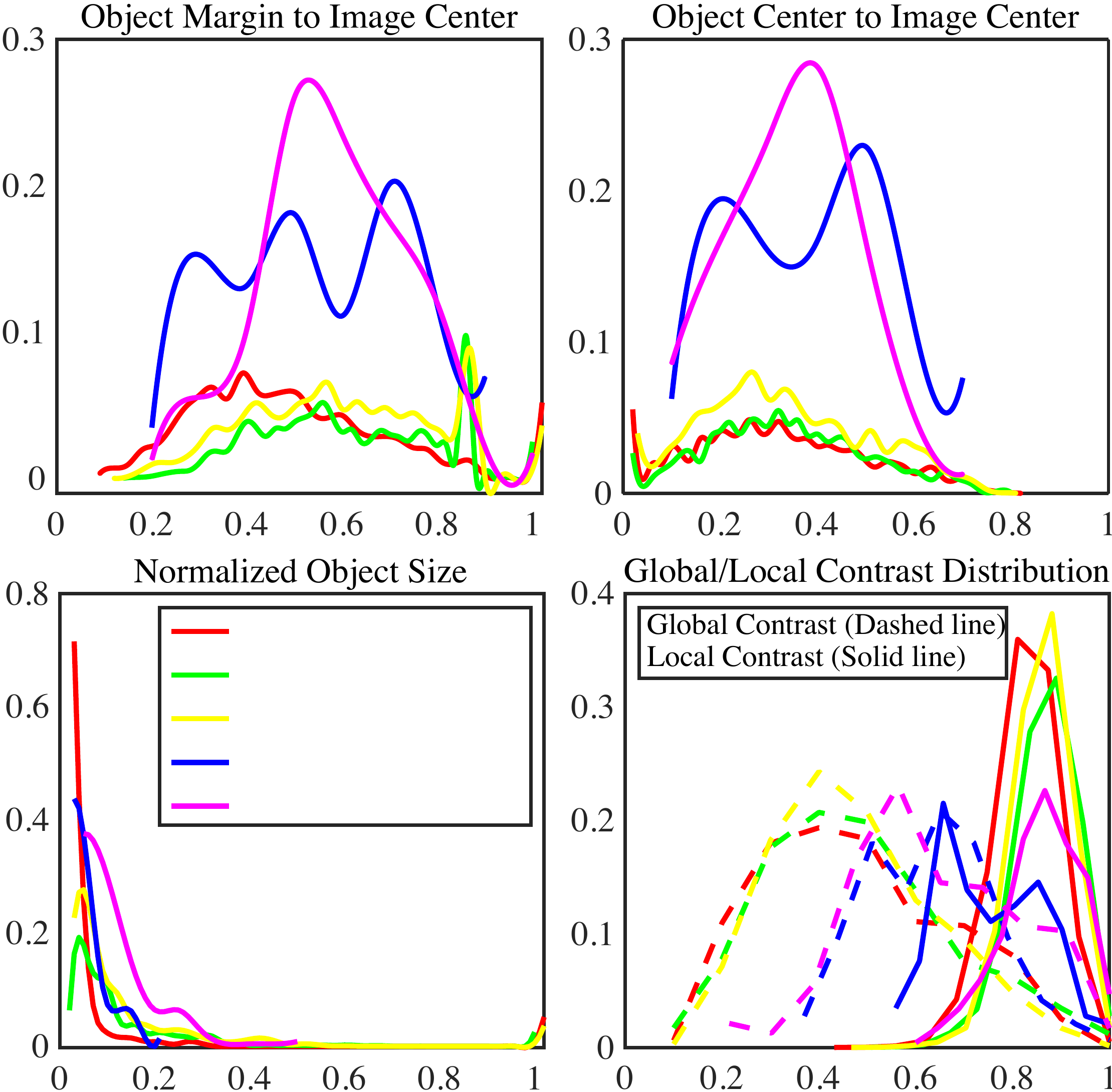}
    \put(-170, 88){\footnotesize \ourdataset-\texttt{Train}}
    \put(-170, 79.25){\footnotesize \ourdataset-\texttt{Easy}}
    \put(-170, 70.5){\footnotesize \ourdataset-\texttt{Hard}}
    \put(-170, 61.75){\footnotesize CVC-300}
    \put(-170, 54){\footnotesize CVC-612}
    \caption{Statistic curves among existing VPS datasets (CVC-300~\&~CVC-612) and our \ourdataset-\texttt{Train}/-\texttt{Easy}/-\texttt{Hard}.
    Note that the horizontal and vertical axes denote the
    frequency and their statistic values, respectively.
    These curves present the diversity of our dataset.
    }
    \label{fig:statistic}
\end{figure}
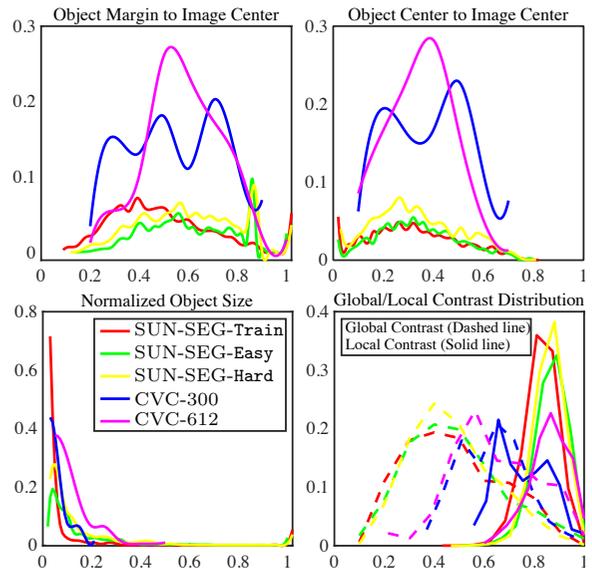

\begin{itemize}
    \item \textit{Center Bias.}
    Unlike general object detection, medical images usually share a higher center bias since the targets are often not in the center of an image.
    To depict the degree of center bias~\cite{fan2021salient}, we compute the average distribution of each dataset's overall ground truth map.
    \figref{fig:center_bias} and \figref{fig:statistic} (top) show that the three sub-datasets of \ourdataset~have lower a center bias than CVC-300 and CVC-612 datasets.
    
    \item \textit{Polyp Size.}
    Colonoscopy is an ego-motion situation instead of shooting moving targets (\ie, stuff and things) through fixed cameras in the general domain.
    As a result, the scale variation of polyps and the irregular movement of the camera causes the different sizes of polyps.
    The polyps partly or even fully disappear in the view.
    \figref{fig:statistic} (bottom-left) shows the comparison of polyp scales at five different VPS datasets.
    
    \item \textit{Global/Local Contrast.}
    To demonstrate how difficult a colon polyp is to identify, in \figref{fig:statistic} (bottom-right), we describe it quantitatively using the global and local contrast strategy~\cite{fan2020rethinking}.
\end{itemize}

\section{VPS Baseline}\label{sec:method}
This section first clarifies the formulation of VPS task in~\secref{sec:task_formulation}.
Then, we describe the details of \ourmodel, including the normalized self-attention block (\secref{sec:NS}), global-to-local learning strategy (\secref{sec:glns_pipeline}), and implementation details (\secref{sec:impl_detail}).

\begin{figure*}[t!]
    \centering
    \begin{overpic}[width=\linewidth]{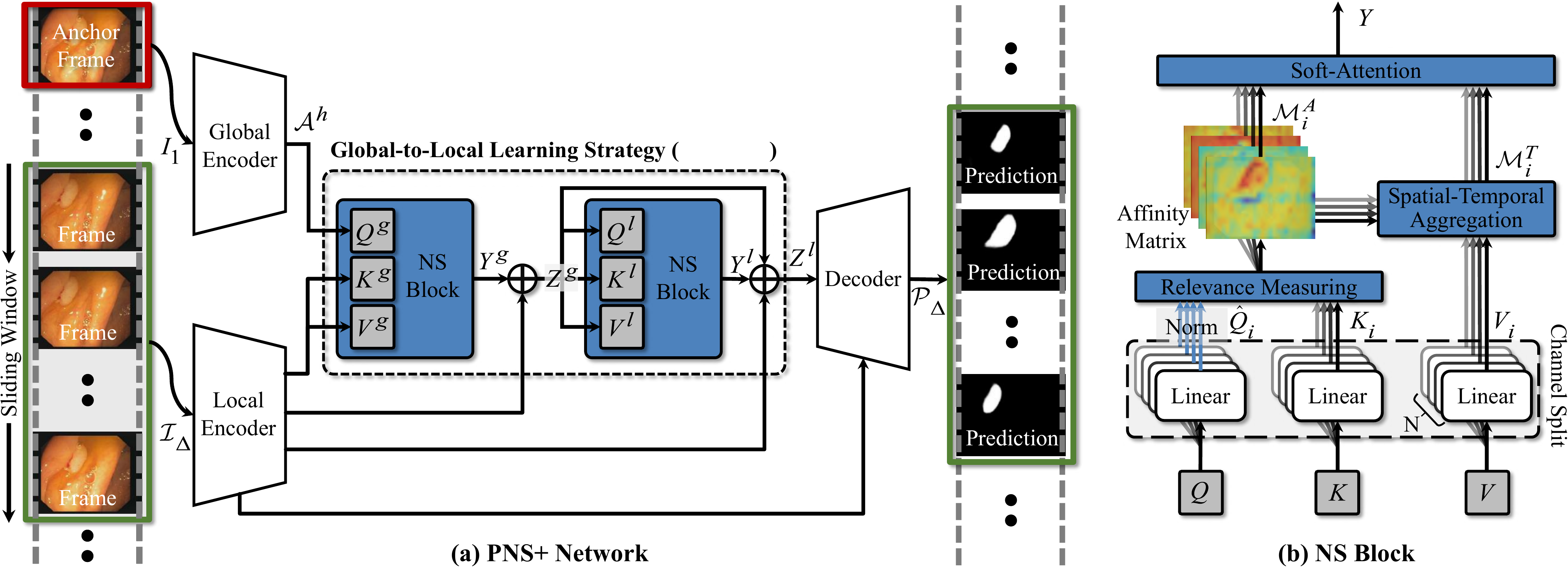}
    \put(18.5, 4.5){$\scriptstyle \{\mathcal{X}^{l}_s\}_{s=t}^{t+\Delta}$}
    \put(18.5, 10.8){$\scriptstyle \{\mathcal{X}^{h}_s\}_{s=t}^{t+\Delta}$}
    \put(31, 8.3){$\scriptstyle \{\mathcal{X}^{h}_s\}_{s=t}^{t+\Delta}$}
    \put(44, 5.8){$\scriptstyle \{\mathcal{X}^{h}_s\}_{s=t}^{t+\Delta}$}
    \put(43.25, 26){\footnotesize \secref{sec:glns_pipeline}}
    \end{overpic}
    \caption{The pipeline of the proposed (a) \ourmodel~network, which is based on (b) the normalized self-attention (NS) block.}\label{fig:main_framework}
\end{figure*}
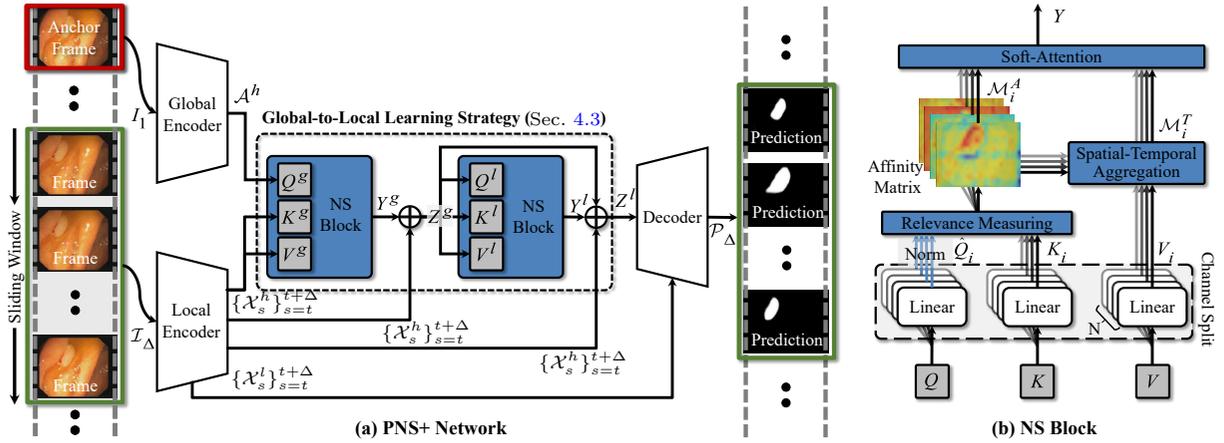

\subsection{Task Formulation}\label{sec:task_formulation}
We mainly focus on the task of video polyp segmentation, which could be defined as a binary-class video object segmentation task, \ie, identifying polyp and non-polyp areas.
Specifically, our goal is to render a model to assign a probability prediction (\ie, a non-binary mask ranging from 0 to 1) for every pixel of each frame.
Besides, we leave other types of tasks for future exploration, such as video polyp detection.

\subsection{Normalized Self-attention Block}\label{sec:NS}
Recently, the self-attention mechanism~\cite{wang2018non} has been widely exploited in many popular computer vision tasks.
Our initial studies found that introducing the original self-attention mechanism to the VPS task does not achieve satisfactory results (high accuracy and speed) due to the multiscale property of polyps that are captured at various shooting angles and speeds.
Directly utilizing the naive self-attention scheme, such as the non-local network \cite{wang2018non}, incurs a high computational cost, limiting the inference speed.
As shown in~\figref{fig:main_framework} (right), we propose a normalized self-attention (NS) block, which is motivated by the fact that dynamically updating the receptive field is important for self-attention-based networks.
The NS block involves five key steps, which are detailed as follows.

\subsubsection{Enhanced Rules}
Motivated by the recent video salient object detection model~\cite{gu2020pyramid}, we utilize three strategies, \ie, \textit{channel split}, \textit{query-dependent}, and \textit{normalization}, to reduce the computational cost and improve the accuracy.

\noindent\textbf{(a) Channel Split Rule.}
Specifically, given three candidate features (\ie, query feature $Q$, key feature $K$, and value feature $V$) with the size of $\mathbb{R}^{T \times H \times W \times C}$, we utilize three linear embedding functions $\theta(\cdot)$, $\phi(\cdot)$, and $g(\cdot)$ to generate the corresponding attention features.
These functions can be implemented by a convolutional layer with a kernel size of $1\times1\times1$~\cite{wang2018non}.
Note that $T$, $H$, $W$ and $C$ denote the number of frames, height, width, and channels of the given feature, respectively.
This rule can be expressed as:
\begin{equation}\label{eqn:channel_split_rule}
    Q_i = \mathcal{F}^G \langle \theta(Q) \rangle, K_i = \mathcal{F}^G \langle \phi(K) \rangle, V_i = \mathcal{F}^G \langle g(V) \rangle,
\end{equation}
where the function $\mathcal{F}^G$ denotes the operation that we split each attention feature into $N$ groups along the channel dimension, resulting in three disparate features: query $Q_i$, key $K_i$, and value $V_i$, where $i = \{ 1, 2, \cdots, N \}$.
Thus, the shape of the above three split features is $\mathbb{R}^{T \times H \times W \times \frac{C}{N}}$.

\noindent\textbf{(b) Query-Dependent Rule.}
To model the spatial-temporal relationship among consecutive frames, we need to measure the similarity between the split query features $\{Q_i\}_{i=1}^N$ and split key features $\{K_i\}_{i=1}^N$.
Inspired by~\cite{gu2020pyramid}, we introduce $N$ relevance measuring (\ie, query-dependent rule) blocks to compute the spatial-temporal affinity matrix for the \textit{constrained neighborhood} of the target pixel.
Rather than computing the response between a query position and a key feature at all positions, as done in~\cite{wang2018non}, the relevance measuring block can capture more relevance regarding the target object within $T$ frames.
More specifically, we get the corresponding constrained neighborhood in $K_i$ for query pixel $\mathbf{X}^q$ of $Q_i$ in position $(x,y,z)$, which can be obtained by a point sampling function $\mathcal{F}^S$.
This is formulated as:
\begin{equation}\label{equ:sampling}
    \mathcal{F}^S \langle \mathbf{X}^q , K_i \rangle = \Sigma_{m=x-kd_i}^{x+kd_i} \Sigma_{n=y-kd_i}^{y+kd_i} \Sigma_{t=1}^{T} K_i(m, n, t),
\end{equation}
where $1 \leq x \leq H$, $1 \leq y \leq W$, and $1 \leq z \leq T$ and $\mathcal{F}^S \langle \mathbf{X}^q , K_i \rangle \in \mathbb{R}^{T(2k+1)^2 \times \frac{C}{N}}$.
Thus, the size of the constrained neighborhood will depend on the various spatial-temporal receptive fields with different kernel sizes $k$, dilation rate $d_i$ at the $i$-th group, and frame number $T$, respectively.

\noindent\textbf{(c) Normalization Rule.}
However, the internal covariate shift problem~\cite{guo2020normalized} exists in the feed-forward of input $Q_i$, incurring that the layer parameters cannot dynamically adapt to the next mini-batch.
Thus, we maintain a fixed distribution for $Q_i$ via:
\begin{equation}
    \hat{Q}_i = \texttt{Norm} (Q_i),
\end{equation}
where $\texttt{Norm}(\cdot)$ is implemented by layer normalization~\cite{ba2016layer} operation along the temporal dimension.

\subsubsection{Relevance Measuring}\label{sec:rel_measuring}
The affinity matrix $\mathcal{M}^A_i$ measures the similarity of target pixels and their surrounding spatial-temporal contents in an adaptive point sampling manner (refer to \eqnref{equ:sampling}).
It is defined as:
\begin{equation}\label{equ:relevance_measuring}
    \mathcal{M}^A_i= \texttt{Softmax}(\frac{\hat{Q}_i \mathcal{F}^S \langle \hat{\mathbf{X}}^q , K_i \rangle^{\top} }{\sqrt{C/N}}),~\text{when}~\hat{\mathbf{X}}^q \in \hat{Q}_i,
\end{equation}
where $\mathcal{M}^A_i \in \mathbb{R}^{THW\times T(2k+1)^2}$.
$\sqrt{C/N}$ is a scaling factor to balance the multi-head attention.

\subsubsection{Spatial-Temporal Aggregation}
Similar to relevance measuring, we also compute the spatial-temporally aggregated features $\mathcal{M}^T_i \in \mathbb{R}^{THW \times \frac{C}{N}}$ within the constrained neighborhood during temporal aggregation.
It is calculated by:
\begin{equation}\label{equ:s_t_aggregating}
    \mathcal{M}^T_i = \mathcal{M}^A_i \mathcal{F}^S \langle \mathbf{X}^a, V_i \rangle,~\text{when}~\mathbf{X}^a \in \mathcal{M}^A_i.
\end{equation}

\subsubsection{Soft-Attention}
We utilize a soft-attention block to synthesize features from the group of affinity matrices $\mathcal{M}^A_i$ and aggregated features $\mathcal{M}^T_i$.
During the synthesis process, relevant spatial-temporal patterns should be enhanced 
while less relevant ones should be suppressed.
We first concatenate a group of affinity matrices $\mathcal{M}^A_i$ along the channel dimension to generate $\mathcal{M}^A$.
The soft-attention map $\mathcal{M}^S$ is computed by:
\begin{equation}
    \mathcal{M}^S \in \mathbb{R}^{THW \times 1} = \texttt{Max}(\mathcal{M}^A),
\end{equation}
where $\mathcal{M}^A \in \mathbb{R}^{THW \times T(2k+1)^2N}$ and the $\texttt{Max}(\cdot)$ function computes the channel-wise maximum value.
We then concatenate a group of the spatial-temporally aggregated features $\mathcal{M}^T_i$ along the channel dimension to generate $\mathcal{M}^T$.

\subsubsection{Normalized Self-attention}
Finally, our normalized self-attention block, \ie, the function $\texttt{NS}(\cdot,\cdot,\cdot)$, is defined as:
\begin{equation}\label{equ:ns_block}
    \begin{split}
    Y \in \mathbb{R}^{T \times {H} \times {W} \times C} &= \texttt{NS}(Q, K, V) \\
    &= (\mathcal{M}^T \mathbf{W}_T) \circledast \mathcal{M}^S,
    \end{split}
\end{equation}
where $\mathbf{W}_T$ is the learnable weight and $\circledast$ denotes the channel-wise Hadamard product.

\subsection{Global-to-Local Learning}\label{sec:glns_pipeline}
\noindent\textbf{Observation.} 
By establishing the non-local connections for the given features, the proposed NS block, as in~\secref{sec:NS}, shows the promising potential for learning short-term spatial-temporal dependencies.
However, this mechanism still struggles in modeling long-term spatial-temporal dependencies due to limited computational resources, \ie, the network can only process a piece of frames within a limited time span.

In contrast to our conference version, PNSNet~\cite{ji2021pnsnet}, we propose a novel global-to-local learning paradigm, which realizes both long-and short-term spatial-temporal propagation at an arbitrary temporal distance, yielding a simple but efficient framework, \ourmodel.
Specifically, it appends a spatial-temporal learning pathway at a global temporal level, naturally introducing long-term dependencies into the network.
We describe this strategy via the following five steps: a global encoder (\secref{sec:global_encoder}), a local encoder (\secref{sec:local_encoder}), the global spatial-temporal modeling (\secref{sec:global_ST_modelling}), the global-to-local propagation (\secref{sec:g2l_propagation}), and the decoder/objectiveness (\secref{sec:decoder_and_objectiveness}).

\subsubsection{Global Encoder}\label{sec:global_encoder}
Our strategy employs the first frame $I_1 \in \mathbb{R}^{H' \times W' \times 3}$ as an anchor (\ie, global reference).
The dependency will be calculated between the anchor frame and the sampled consecutive frames within a sliding window.
Following PraNet~\cite{fan2020pra}, we use the same backbone, Res2Net-50~\cite{pami20Res2net}, to extract the feature at the \texttt{conv4\_6} layer.
To alleviate the computational burden, we adopt an RFB-like~\cite{liu2018receptive} module to reduce the channel dimension of extracted feature and generate the anchor feature $\mathcal{A}^h \in \mathbb{R}^{H^{h} \times W^{h} \times C^{h}}$.

\subsubsection{Local Encoder}\label{sec:local_encoder}
The local encoder takes a piece of consecutive frames $\mathcal{I}_\Delta$$=$$\{ I_s \}_{s=t}^{t+\Delta} \in \mathbb{R}^{H' \times W' \times 3}~(t$$>$$1)$ from a sliding window as input.
Similar to the global encoder, we leverage Res2Net-50 backbone to extract two groups of short-term features from the \texttt{conv3\_4} and \texttt{conv4\_6} layers and use channel reduction to generate the low-level $\{\mathcal{X}^{l}_s\}_{s=t}^{t+\Delta}$$\in$$\mathbb{R}^{H^{l} \times W^{l} \times C^{l}}$ and high-level $\{\mathcal{X}^{h}_s\}_{s=t}^{t+\Delta}$$\in$$\mathbb{R}^{H^{h} \times W^{h} \times C^{h}}$ short-term features.
We set $H^{l}=\frac{H'}{4}$, $W^{l}=\frac{W'}{4}$, $C^{l}=24$, $H^{h}=\frac{H'}{8}$, $W^{h}=\frac{W'}{8}$, and $C^{h}=32$ as the default implementation.

\subsubsection{Global Spatial-Temporal Modeling}\label{sec:global_ST_modelling}
As shown in~\figref{fig:main_framework}, we leverage the first NS block to model the long-term relationship at an arbitrary temporal distance, which requires a four-dimensional temporal feature as input; therefore we have: 
\begin{equation}
    \begin{aligned}
    \tilde{\mathcal{X}}^h \in \mathbb{R}^{\Delta \times H^{h} \times W^{h} \times C^{h}} &\Leftarrow \{\mathcal{X}^{h}_s\}_{s=t}^{t+\Delta} \in \mathbb{R}^{H^{h} \times W^{h} \times C^{h}}, \\
    \tilde{\mathcal{A}}^h \in \mathbb{R}^{1 \times H^{h} \times W^{h} \times C^{h}} &\Leftarrow \mathcal{A}^h \in \mathbb{R}^{H^{h} \times W^{h} \times C^{h}},
    \end{aligned}
\end{equation}
where $\Leftarrow$ denotes reshaping the candidate features into the temporal form to yield a four-dimensional tensor.
Then, as for the first NS block formulated in \eqnref{equ:ns_block}, we employ the anchor feature as a query entry (\ie, $Q^g=\tilde{\mathcal{A}}^h$) and the high-level short-term feature as the key and value entries (\ie, $K^g$$=$$\tilde{\mathcal{X}}^h$ \& $V^g$$=$$\tilde{\mathcal{X}}^h$).
Intuitively, we aim to build the pixel-wise similarities between the anchor and high-level short-term features, which could be viewed as the modeling of global spatial-temporal dependencies.
It is defined as:
\begin{equation}
    Z^g \in \mathbb{R}^{\Delta \times H^{h} \times W^{h} \times C^{h}}= \texttt{NS}(\tilde{\mathcal{A}}^h, \tilde{\mathcal{X}}^h,\tilde{\mathcal{X}}^h) \oplus \tilde{\mathcal{X}}^h,
\end{equation}
where $\oplus$ denotes the element-wise addition of residual operation~\cite{he2016deep}.
This operation provides better convergence stability of interior gradient propagation within the first NS block, allowing it to easily be plugged into the pre-trained networks.

\subsubsection{Global-to-Local Propagation}\label{sec:g2l_propagation}
Furthermore, we desire to propagate the long-term dependency $Z^g$ into a local neighborhood (\ie, frames in a sliding window).
Thus, we serve $Z^g$ as the input entries of the second NS block as in \eqnref{equ:ns_block}, \ie, query $Q^l=Z^g$, key $K^l=Z^g$, and value $V^l=Z^g$.
We have:
\begin{equation}
    Z^l = \texttt{NS}(Z^g,Z^g,Z^g) \oplus Z^g \oplus \tilde{\mathcal{X}}^h.
\end{equation}
In this way, the introduced two residual connections 
can maintain the interior gradient stability (\ie, $\oplus Z^g$) and exterior gradient stability (\ie, $\oplus \tilde{\mathcal{X}}^h$) of the second NS block.

\subsubsection{Decoder and Objectiveness}\label{sec:decoder_and_objectiveness}
Finally, we combine the low-level short-term feature $\mathcal{X}^l_s$ from the local encoder and the spatial-temporal feature $Z^l$ from the second NS block with a two-stage UNet-alike decoder $\mathcal{F}^D$.
Before the combination, we recover the feature $Z^l$ back to the spatial form, \ie, $\{Z^l_s\}_{s=t}^{t+\Delta}$.
The prediction from the decoder is computed with:
\begin{equation}
    \mathcal{P}_\Delta = \{P_s\}_{s=t}^{t+\Delta} = \mathcal{F}^D \langle \{\mathcal{X}^{l}_s\}_{s=t}^{t+\Delta}, \{Z^l_s\}_{s=t}^{t+\Delta} \rangle.
\end{equation}
To this end, given a prediction $P_s$ and corresponding ground-truth (GT) $G_s$ at timestamp $s$, we utilize a \textit{binary cross-entropy} loss for optimization, which is formulated as:
\begin{equation}
    \mathcal{L}_{bce} = - \sum [ G_s \log( P_s ) + ( 1-G_s ) \log( 1-P_s) ) ].
\end{equation}

\subsection{Implementation Details}\label{sec:impl_detail}
\subsubsection{Datasets}\label{sec:data_split}
We split $40\%$ \ourdataset~data for training, \ie, \ourdataset-\texttt{Train} with 112 clips ($19,544$ frames).
The rest data are all used for testing, including \ourdataset-\texttt{Easy} with 119 clips ($17,070$ frames) and \ourdataset-\texttt{Hard} with 54 clips ($12,522$ frames) according to difficulty levels in each pathological category.
Specifically, two colonoscopy scenarios (\ie, seen and unseen\footnote{Seen denotes that the samples in the testing dataset are from the same case in the training set, whereas the unseen indicates that the scenario do not exist in the training set.}) are included in the above two testing dataset: \ourdataset-\texttt{Easy} (seen: 33 clips \& unseen: 86 clips) and \ourdataset-\texttt{Hard} (seen: 17 clips \& unseen: 37 clips) for more fine-grained experimental analyses.

\subsubsection{Training Details}\label{sec:training_settings}
We train our model using the \ourdataset-\texttt{Train} dataset on the server platform equipped with an Intel Xeon (R) CPU E5-2690v4$\times$24 and four NVIDIA Tesla V100 GPUs with 16 GB memory per one.
The ImageNet pre-trained weights of Res2Net-50~\cite{pami20Res2net} are loaded before training, and other newly-added layers are with Kaiming initialization.
We set the batch size to $24$, which takes about 5 hours to reach convergence after 15 epochs.
For each mini-batch of data, we select the first frame of a video clip as an anchor, randomly sample five consecutive frames ($\Delta$$=$$5$) from the same clip, and resize them to 256$\times$448.
The Adam optimizer's initial learning rate and weight decay are set to 3e-4 and 1e-4, respectively.
We set the number of attention groups to $N$$=$$4$ as default.
For the first NS block, we set kernel size $k$$=$$3$ and dilation rate $d_i$$=$$\{3,4,3,4\}$ to capture more long-term representations with a larger receptive field.
For the second one, we set kernel size $k$$=$$3$ and reduce dilation rate $d_i$$=$$\{1,2,1,2\}$ to mainly focus on short-term relationships.

\subsubsection{Inference Stage}
We evaluate \ourmodel~on \ourdataset-\texttt{Easy} and \ourdataset-\texttt{Hard} with both seen and unseen scenarios.
Similar to the training phase, during inference, we select the first frame as an anchor, sample five video frames ($\Delta$$=$$5$) from a video clip, and resize them to 256$\times$448.
For the final prediction, we use the network's output $\mathcal{P}_\Delta$ followed by a \textit{Sigmoid} function.
The proposed \ourmodel~achieves a super real-time inference speed of 170fps on a single V100 GPU without any heuristic post-processing techniques, such as DenseCRF~\cite{krahenbuhl2011efficient}.

\section{VPS Benchmark}\label{sec:vps_benchmark}

\subsection{Evaluation Protocols}\label{sec:evaluation_protocols}

\subsubsection{Competitors} 
We elaborately select eight typical video-based object/polyp segmentation methods, including COSNet~\cite{lu2019see}, MAT~\cite{zhou2020matnet}, PCSA~\cite{gu2020pyramid}, 2/3D~\cite{puyal2020endoscopic}, AMD~\cite{liu2021emergence}, DCF~\cite{zhang2021dynamic}, FSNet~\cite{ji2021full}, and PNSNet~\cite{ji2021pnsnet}.
We also add five image-based object/polyp segmentation methods to validate the effectiveness of per-frame prediction ability, including UNet~\cite{ronneberger2015u}, UNet++~\cite{zhou2018unetplus}, ACSNet~\cite{zhang2020adaptive}, PraNet~\cite{fan2020pra}, and SANet~\cite{wei2021shallow}.
For a fair comparison, all the competitors utilize the same dataset as our \ourmodel~and reach the convergence under their default training settings.
Of note, this paper only focuses on the positive cases in our \ourdataset~dataset and left negative cases (no polyp) for future work.

\subsubsection{Evaluation Metrics} 
To provide deeper insight into the model performance, we use the following six different metrics for model evaluation between prediction $P_s$ and ground-truth $G_s$ at timestamp $s$, including:
(a) Dice coefficient (${\rm Dice} = \frac{ 2 \times \vert P_s \cap G_s \vert}{\vert P_s \cup G_s \vert}$) measures the similarity between prediction and ground truth mask and penalizes for the false-positive/-negative predictions.
The operators $\cap$, $\cup$, and $\vert \cdot \vert$ denote the intersection, union, and the number of pixels in an area, respectively.
(b) Pixel-wise sensitivity (${\rm Sen} = \frac{ \vert P_s \cap G_s \vert}{\vert G_s \vert}$) is used to evaluate the true positive prediction of overall lesion areas.
Since the goal of colonoscopy is to screen the polyps with a low polyp missing rate, people who have the polyps should be highly likely to be identified.
As a result, penalizing the false-negative prediction can be done by adopting sensitivity which refers to the method's ability to correctly detect polyps.
(c) Being the harmonic mean of precision and recall that is weighted by $\beta$, F-measure~\cite{achanta2009frequency} ($F_{\beta} = \frac{(1+\beta^{2}) \times {\rm Prc} \times {\rm Rcl}}{\beta^{2} \times ( {\rm Prc} + {\rm Rcl} )}$)
is widely used in measuring binary masks by combining precision (${\rm Prc} = \frac{\vert P_{s} \cap G_{s} \vert}{\vert P_{s}\vert}$) and recall (${\rm Rcl} = \frac{\vert P_{s} \cap G_{s} \vert}{\vert G_{s}\vert}$) for more comprehensive evaluation.
(d) Suggested by~\cite{fan2021cognitive,cheng2021structure}, weighted F-measure~\cite{margolin2014evaluate} ($F_{\beta}^{w} = \frac{(1+\beta^{2}) \times {\rm Prc}^{w} \times {\rm Rcl}^{w}}{\beta^{2} \times ( {\rm Prc}^{w} + {\rm Rcl}^{w} )}$):
amend the ``Equal-importance flaw'' in Dice and $F_{\beta}$, providing more reliable evaluation results.
Following~\cite{borji2015salient}, we set the factor $\beta^{2}$ of $F_{\beta}$ and $F_{\beta}^{w}$ as 0.3 and 1, respectively.
(e) Different from the above pixel-wise metrics, structure measure~\cite{fan2017structure} ($\mathcal{S}_{\alpha} = \alpha \times \mathcal{S}_{o}(P_{s}, G_{s}) + (1-\alpha) \times \mathcal{S}_{r}(P_{s}, G_{s})$):
is used to measure the structural similarity at object-aware $\mathcal{S}_{o}$ and region-aware $\mathcal{S}_{r}$, respectively.
We use the factor $\alpha$ = 0.5 as default.
(f) Fan~\etal~proposed a human visual perception-based metric, enhanced-alignment measure~\cite{fan2018enhanced}: $E_{\phi} = \frac{1}{W \times H} \sum_{x}^{W} \sum_{y}^{H} \phi (P_{s}(x,y), G_{s}(x,y))$, where $\phi$ is the enhanced-alignment matrix.
$W$ and $H$ are the width and height of ground-truth $G_{s}$.
This metric is inherently suitable for assessing polyps' heterogeneous location and shape in colonoscopy.

\begin{table}[t!]
    \centering 
    \footnotesize
    \renewcommand{\arraystretch}{1.2}
    \setlength\tabcolsep{1pt}
    \caption{Quantitative comparison on two testing sub-datasets with seen colonoscopy scenarios.}\label{tab:ModelScore_Seen}
    \begin{tabular}{r||cccc|cccc} 
    \hline
    & \multicolumn{4}{c|}{\tabincell{c}{\ourdataset-\texttt{Easy} (Seen)}} &\multicolumn{4}{c}{\tabincell{c}{\ourdataset-\texttt{Hard} (Seen)}} \\
    Model & $\mathcal{S}_{\alpha}$ & $E_\phi^{mn}$ & $F_\beta^w$ & Dice
    & $\mathcal{S}_{\alpha}$ & $E_\phi^{mn}$ & $F_\beta^w$ & Dice\\
    \hline
    COSNet & 0.845 & 0.836 & 0.727 & 0.804 & 0.785 & 0.772 & 0.626 & 0.725 \\
    MAT & 0.879 & 0.861 & 0.731 & 0.833 & 0.840 & 0.821 & 0.652 & 0.776 \\
    PCSA & 0.852 & 0.835 & 0.681 & 0.779 & 0.772 & 0.759 & 0.566 & 0.679 \\
    2/3D & 0.895 & 0.909 & 0.819 & 0.856 & 0.849 & 0.868 & 0.753 & 0.809 \\
    AMD & 0.471 & 0.526 & 0.114 & 0.245 & 0.480 & 0.536 & 0.115 & 0.231 \\
    DCF & 0.572 & 0.591 & 0.357 & 0.398 & 0.603 & 0.602 & 0.385 & 0.443 \\
    FSNet & 0.890 & 0.895 & 0.818 & 0.873 & 0.848 & 0.859 & 0.755 & 0.828 \\
    PNSNet & 0.906 & 0.910 & 0.836 & 0.861 & 0.870 & 0.892 & 0.787 & 0.823 \\
    \hline
    \rowcolor{mygray}
    \textbf{\ourmodel} & \textbf{0.917} & \textbf{0.924} & \textbf{0.848} & \textbf{0.888} & \textbf{0.887} & \textbf{0.929} & \textbf{0.806} & \textbf{0.855} \\
    \hline
    \end{tabular}
\end{table}

As mentioned in~\secref{sec:task_formulation}, the models generate continuous floating predictions; thus, we need to threshold the floating value into binary ones ranging from 0 to 255.
Specifically, we provide the maximum value of Dice and mean value of $E_{\phi}$, $F_{\beta}$, and Sen under different thresholds for the binary metrics.
Furthermore, we obtain the video-level score by averaging the evaluated results per image at a video clip.
Then, we take the average video-level scores as the performance on the whole dataset.
One-key evaluation toolbox is available at \url{https://github.com/GewelsJI/VPS/tree/main/eval}.

\subsection{Quantitative Comparison}\label{sec:benchmark}
Based on the protocols mentioned in~\secref{sec:evaluation_protocols}, we conduct a comprehensive VPS benchmark on two testing sub-datasets (\ie, \ourdataset-\texttt{Easy} and \ourdataset-\texttt{Hard}), which include the following three aspects.

\subsubsection{Learning Ability}
Notably, the image-based models are trained and inferred frame-by-frame.
To better unveil the spatial-temporal learning ability on the colonoscopy videos, we conduct two groups of experiments to validate video-based competitors' ability on two seen sub-datasets.
For these sub-datasets shown in \tabref{tab:ModelScore_Seen}, our \ourmodel~also outperforms top-1 video-based approaches, \eg, Dice score on \ourdataset-\texttt{Easy} (Seen): PNSNet (0.861) \textit{vs.} \ourmodel (0.888) and $F^{mn}_\phi$ score on \ourdataset-\texttt{Hard} (Seen): PNSNet (0.892) \textit{vs.} \ourmodel (0.929).
The above results suggest that our model has a strong learning ability to accurately segment polyps.

\subsubsection{Generalization Capability}
To validate the model's generalizability, we conduct the experiments on two testing sub-datasets with unseen colonoscopy scenarios.
As shown in~\tabref{tab:ModelScore}, we present the performance comparison with the other latest image- and video-based competitors in six metrics.
It shows that our \ourmodel~achieves significant improvements by a large margin in comparison with top image-and video-based approaches, \eg, Dice score on \ourdataset-\texttt{Easy} (Unseen): ACSNet (0.713) \textit{vs.} 2/3D (0.722) \textit{vs.} \ourmodel~(0.756) and $F^w_\beta$ score on \ourdataset-\texttt{Hard} (Unseen): ACSNet (0.636) \textit{vs.} 2/3D (0.634) \textit{vs.} \ourmodel~(0.653).
Interestingly, we observe that PNSNet drops dramatically on two unseen datasets, which is a side show of better generalizability attributed to our newly-proposed global-to-local learning strategy, especially on a clip with a larger time span.

\begin{table*}[t!]
    \centering
    \footnotesize
    \renewcommand{\arraystretch}{1.0}
    \setlength\tabcolsep{3pt}
    \caption{Quantitative comparison of two testing sub-datasets with unseen colonoscopy scenarios.
    `R/T' means to retrain the private model using the code provided by the author.
    The best values are highlighted in \textbf{bold}.}
    \label{tab:ModelScore}
    \begin{tabular}{c|rcr||cccccc|cccccc} 
    \hline
    & && &\multicolumn{6}{c|}{\tabincell{c}{\ourdataset-\texttt{Easy} (Unseen)}} &\multicolumn{6}{c}{\tabincell{c}{\ourdataset-\texttt{Hard} (Unseen)}}\\
    & Model & Publish & Code & $\mathcal{S}_{\alpha}$ & $E_\phi^{mn}$ & $F_\beta^w$ & $F_\beta^{mn}$  & Dice & Sen
    & $\mathcal{S}_{\alpha}$ & $E_\phi^{mn}$ & $F_\beta^w$ & $F_\beta^{mn}$  & Dice & Sen\\
    \hline 
    \multirow{5}{*}{\rotatebox[origin=C]{90}{\textbf{IMAGE}}}
    & UNet~\cite{ronneberger2015u} & MICCAI$_{15}$& \href{Link}{https://github.com/4uiiurz1/pytorch-nested-unet}& 0.669 & 0.677 & 0.459 & 0.528 & 0.530 & 0.420 & 0.670 & 0.679 & 0.457 & 0.527 & 0.542 & 0.429 \\
    & UNet++~\cite{zhou2018unetplus} & TMI$_{18}$& \href{Link}{https://github.com/MrGiovanni/UNetPlusPlus}& 0.684 & 0.687 & 0.491 & 0.553 & 0.559 & 0.457 & 0.685 & 0.697 & 0.480 & 0.544 & 0.554 & 0.467 \\
    & ACSNet~\cite{zhang2020adaptive} & MICCAI$_{20}$&\href{Link}{https://github.com/ReaFly/ACSNet}& 0.782 & 0.779 & 0.642 & 0.688 & 0.713 & 0.601 & 0.783 & 0.787 & 0.636 & 0.684 & 0.708 & 0.618 \\
    & PraNet~\cite{fan2020pra} & MICCAI$_{20}$&\href{Link}{https://github.com/DengPingFan/PraNet}& 0.733 & 0.753 & 0.572 & 0.632 & 0.621 & 0.524 & 0.717 & 0.735 & 0.544 & 0.607 & 0.598 & 0.512 \\
    & SANet~\cite{wei2021shallow} & MICCAI$_{21}$&\href{Link}{https://github.com/weijun88/SANet}& 0.720 & 0.745 & 0.566 & 0.634 & 0.649 & 0.521 & 0.706 & 0.743 & 0.526 & 0.580 & 0.598 & 0.505 \\
    \hline
    \multirow{8}{*}{\rotatebox[origin=C]{90}{\textbf{VIDEO}}} 
    & COSNet~\cite{lu2019see} & TPAMI$_{19}$&\href{Link}{https://github.com/carrierlxk/COSNet}& 0.654 & 0.600 & 0.431 & 0.496 & 0.596 & 0.359 & 0.670 & 0.627 & 0.443 & 0.506 & 0.606 & 0.380 \\
    & MAT~\cite{zhou2020matnet} & TIP$_{20}$&\href{Link}{https://github.com/tfzhou/MATNet}& 0.770 & 0.737 & 0.575 & 0.641 & 0.710 & 0.542 & 0.785 & 0.755 & 0.578 & 0.645 & 0.712 & 0.579 \\
    & PCSA~\cite{gu2020pyramid} & AAAI$_{20}$&\href{Link}{https://github.com/guyuchao/PyramidCSA}& 0.680 & 0.660 & 0.451 & 0.519 & 0.592 & 0.398 & 0.682 & 0.660 & 0.442 & 0.510 & 0.584 & 0.415 \\
    & 2/3D~\cite{puyal2020endoscopic} & MICCAI$_{20}$& R/T& 0.786 & 0.777 & 0.652 & 0.708 & 0.722 & 0.603 & 0.786 & 0.775 & 0.634 & 0.688 & 0.706 & 0.607 \\
    & AMD~\cite{liu2021emergence} & NeurIPS$_{21}$&\href{Link}{https://github.com/rt219/the-emergence-of-objectness}& 0.474 & 0.533 & 0.133 & 0.146 & 0.266 & 0.222 & 0.472 & 0.527 & 0.128 & 0.141 & 0.252 & 0.213 \\
    & DCF~\cite{zhang2021dynamic} & ICCV$_{21}$&\href{Link}{https://github.com/Roudgers/DCFNet}& 0.523 & 0.514 & 0.270 & 0.312 & 0.325 & 0.340 & 0.514 & 0.522 & 0.263 & 0.303 & 0.317 & 0.364 \\
    & FSNet~\cite{ji2021full} & ICCV$_{21}$&\href{Link}{https://github.com/GewelsJI/FSNet}& 0.725 & 0.695 & 0.551 & 0.630 & 0.702 & 0.493 & 0.724 & 0.694 & 0.541 & 0.611 & 0.699 & 0.491 \\
    & PNSNet~\cite{ji2021pnsnet} & MICCAI$_{21}$&\href{Link}{https://github.com/GewelsJI/PNS-Net}& 0.767 & 0.744 & 0.616 & 0.664 & 0.676 & 0.574 & 0.767 & 0.755 & 0.609 & 0.656 & 0.675 & 0.579 \\
    \hline
    \rowcolor{mygray}
    & \textbf{\ourmodel} & \textbf{OURS$_{22}$}&\href{Link}{https://github.com/GewelsJI/VPS} & \textbf{0.806} & \textbf{0.798} & \textbf{0.676} & \textbf{0.730} & \textbf{0.756} & \textbf{0.630}& \textbf{0.797} & \textbf{0.793} & \textbf{0.653} & \textbf{0.709} & \textbf{0.737} & \textbf{0.623} \\
    \hline
    \end{tabular}
\end{table*}

\begin{table*}[t!]
    \centering
    \footnotesize
    \renewcommand{\arraystretch}{1.0}
    \setlength\tabcolsep{0.2pt}
    \caption{Visual attributes-based performance on \ourdataset-\texttt{Easy}/-\texttt{Hard} (Unseen) in terms of structure measure ($\mathcal{S}_\alpha$) score.}
    \label{tab:AttributeScore}
    \begin{tabular}{r||cccccccccc|cccccccccc} 
    \hline
    & \multicolumn{10}{c|}{\tabincell{c}{\ourdataset-\texttt{Easy} (Unseen)}} &\multicolumn{10}{c}{\tabincell{c}{\ourdataset-\texttt{Hard} (Unseen)}}\\
    & \textbf{SI} & \textbf{IB} & \textbf{HO} & \textbf{GH} & \textbf{FM} & \textbf{SO} & \textbf{LO} & \textbf{OC} & \textbf{OV} & \textbf{SV} & \textbf{SI} & \textbf{IB} & \textbf{HO} & \textbf{GH} & \textbf{FM} & \textbf{SO} & \textbf{LO} & \textbf{OC} & \textbf{OV} & \textbf{SV} \\
    \hline
    UNet & 0.675 & 0.548 & 0.768 & 0.715 & 0.633 & 0.593 & 0.648 & 0.670 & 0.643 & 0.620 & 0.618 & 0.619 & 0.663 & 0.676 & 0.713 & 0.689 & 0.633 & 0.658 & 0.659 & 0.658 \\
    UNet++ & 0.701 & 0.542 & 0.782 & 0.739 & 0.647 & 0.591 & 0.678 & 0.683 & 0.665 & 0.617 & 0.654 & 0.604 & 0.665 & 0.696 & 0.714 & 0.681 & 0.660 & 0.676 & 0.677 & 0.678 \\
    ACSNet & 0.789 & 0.612 & 0.896 & 0.820 & 0.704 & 0.663 & 0.787 & 0.770 & 0.759 & 0.705 & 0.770 & 0.681 & 0.828 & 0.795 & 0.817 & 0.738 & 0.810 & \textbf{0.828}  & 0.806 & 0.759 \\
    PraNet & 0.745 & 0.585 & 0.821 & 0.772 & 0.673 & 0.611 & 0.722 & 0.722 & 0.703 & 0.653 & 0.673 & 0.635 & 0.725 & 0.720 & 0.755 & 0.691 & 0.666 & 0.714 & 0.708 & 0.703 \\
    SANet & 0.724 & 0.582 & 0.854 & 0.760 & 0.676 & 0.615 & 0.703 & 0.701 & 0.711 & 0.680 & 0.658 & 0.565 & 0.738 & 0.709 & 0.760 & 0.692 & 0.733 & 0.729 & 0.727 & 0.693 \\
    \hline
    COSNet & 0.663 & 0.531 & 0.786 & 0.684 & 0.610 & 0.549 & 0.637 & 0.648 & 0.613 & 0.617 & 0.641 & 0.593 & 0.727 & 0.668 & 0.690 & 0.637 & 0.694 & 0.707 & 0.666 & 0.625 \\
    MAT & 0.772 & 0.664 & 0.873 & 0.789 & 0.706 & \textbf{0.691}  & 0.755 & 0.738 & 0.746 & 0.715 & \textbf{0.772}  & 0.701 & 0.801 & 0.776 & 0.782 & 0.780 & 0.791 & 0.795 & 0.789 & 0.750 \\
    PCSA & 0.676 & 0.563 & 0.759 & 0.708 & 0.628 & 0.610 & 0.634 & 0.662 & 0.656 & 0.616 & 0.656 & 0.591 & 0.692 & 0.683 & 0.706 & 0.671 & 0.612 & 0.677 & 0.665 & 0.663 \\
    2/3D & 0.809 & 0.625 & \textbf{0.899}  & 0.835 & 0.728 & 0.667 & \textbf{0.820}  & \textbf{0.783}  & 0.778 & 0.719 & 0.768 & 0.662 & \textbf{0.865}  & 0.784 & 0.797 & 0.737 & \textbf{0.853}  & 0.827 & \textbf{0.808}  & 0.765 \\
    AMD & 0.476 & 0.461 & 0.471 & 0.481 & 0.484 & 0.466 & 0.447 & 0.467 & 0.442 & 0.498 & 0.471 & 0.468 & 0.447 & 0.473 & 0.468 & 0.469 & 0.453 & 0.487 & 0.462 & 0.481 \\
    DCF & 0.465 & 0.485 & 0.479 & 0.505 & 0.541 & 0.495 & 0.362 & 0.484 & 0.492 & 0.495 & 0.441 & 0.508 & 0.422 & 0.498 & 0.587 & 0.556 & 0.351 & 0.470 & 0.494 & 0.540 \\
    FSNet & 0.719 & 0.603 & 0.810 & 0.752 & 0.694 & 0.632 & 0.686 & 0.711 & 0.691 & 0.665 & 0.662 & 0.648 & 0.743 & 0.713 & 0.774 & 0.723 & 0.701 & 0.728 & 0.728 & 0.694 \\
    PNSNet & 0.789 & 0.592 & 0.871 & 0.820 & 0.723 & 0.619 & 0.768 & 0.749 & 0.751 & 0.705 & 0.746 & 0.631 & 0.803 & 0.780 & 0.778 & 0.743 & 0.805 & 0.790 & 0.794 & 0.758 \\
    \hline
    \rowcolor{mygray}
    \textbf{\ourmodel}& \textbf{0.819} & \textbf{0.667} & 0.883 & \textbf{0.844} & \textbf{0.738} & 0.690 & 0.796 & 0.782 & \textbf{0.798} & \textbf{0.734} & 0.770 & \textbf{0.703} & 0.817 & \textbf{0.801} & \textbf{0.823} & \textbf{0.793} & 0.792 & 0.808 & 0.807 & \textbf{0.795} \\
    \hline
    \end{tabular}
\end{table*}

\subsubsection{Attribute-based Performance}
Finally, we analyze the visual attribute-based comparison presented in~\tabref{tab:Attr}.
In terms of $\mathcal{S}_\alpha$ score, \tabref{tab:AttributeScore} unveils that our \ourmodel~consistently outperforms other rivals on four attributes (\ie, IB, GH, FM, and SV).
More specifically, as shown in~\tabref{tab:AttributeScore}, most methods can not address the VPS tasks with IB attribute since the colon polyps always have fuzzy boundaries.
In contrast, \ourmodel~achieves the best score ($\mathcal{S}_{\alpha}=0.667$) on this challenging IB attribute of \ourdataset-\texttt{Easy} (Unseen).
This discovery is also consistent with the results shown in \figref{fig:front_figure}.
Similarly, the SO attributes also present the lower scores (\eg, \ourdataset-\texttt{Easy} (Unseen): $\mathcal{S}_\alpha$$=$$0.667$), which indicates these two attributes are the most challenging issues in colonoscopy.
On the contrary, HO and LO attribute consistently sustain higher scores than other attributes, making polyp easier to detect.
This phenomenon meets our expectations since the less distribution bias for these relatively easy scenarios.
We refer the readers to~\secref{sec:failure_case} for a more visualized analysis of challenging cases.

\subsection{Qualitative Comparison}
As shown in~\figref{fig:front_figure}, we present visual results on three video clips of four typical models (\ie, PNSNet, 2/3D, MAT, ACSNet) and our \ourmodel.
In the last four rows, the competitors fail to generate complete segmentation results for the polyps that share the same camouflaged texture with the background.
In contrast, in the $3^{rd}$ row, our model can accurately locate and segment polyps in a challenging situation, \ie, polyps with different sizes and homogeneous textures.

\subsection{Ablation Studies}\label{sec:ablation}
To validate the effectiveness of our core designs, we conduct extensive ablation studies and summarize the results in~\tabref{tab:AblationScore}.

\subsubsection{Contribution of Base Network}
We initialize an UNet-like variant \#01 via leveraging the Res2Net-50~\cite{pami20Res2net} backbone, which can be viewed as an image-based approach to generate per frame predictions.
We observe that \textbf{\#OUR} significantly improves the performance ($\mathcal{S}_\alpha$: +7.7\%) on \ourdataset-\texttt{Easy} (Unseen).

\begin{table*}[t!]
    \centering
    \footnotesize
    \renewcommand{\arraystretch}{1.0}
    \setlength\tabcolsep{6.pt}
    \caption{Ablation studies for the core designs of the proposed \ourmodel.
    See~\secref{sec:ablation} for the detailed analyses.}\label{tab:AblationScore}
    \begin{tabular}{c | ccccc || cccc | cccc} 
    \hline 
    & \multicolumn{5}{c||}{\tabincell{c}{\textbf{VARIANTS}}}  
    & \multicolumn{4}{c|}{\tabincell{c}{\ourdataset-\texttt{Easy} (Unseen)}}
    & \multicolumn{4}{c}{\tabincell{c}{\ourdataset-\texttt{Hard} (Unseen)}} \\ 
    No. & Base & $N$ & Soft &Norm & Strategy
    & $\mathcal{S}_{\alpha}$  & $E_\phi^{mn}$  & $F_\beta^w$  & Dice
    & $\mathcal{S}_{\alpha}$  & $E_\phi^{mn}$  & $F_\beta^w$  & Dice \\ 
    \hline
    \#01 & \checkmark &- &- &- &- & 0.729 & 0.718 & 0.571 & 0.616 & 0.726 & 0.720 & 0.559 & 0.603 \\
    \hline
    \#02 & \checkmark & 1 &\checkmark &\checkmark &L & 0.782 & 0.766 & 0.631 & 0.722 & 0.783 & 0.775 & 0.629 & 0.715 \\
    \#03 & \checkmark & 2 &\checkmark &\checkmark &L & 0.773 & 0.760 & 0.625 & 0.720 & 0.785 & 0.784 & 0.631 & 0.719 \\
    \#04 & \checkmark & 4 &\checkmark &\checkmark &L & 0.786 & 0.777 & 0.651 & 0.741 & 0.792 & 0.789 & 0.649 & 0.735 \\
    \#05 & \checkmark & 8 &\checkmark &\checkmark &L & 0.774 & 0.762 & 0.627 & 0.724 & 0.775 & 0.774 & 0.619 & 0.708 \\
    \hline
    \#06 & \checkmark & 4 &- & \checkmark &L & 0.782 & 0.775 & 0.639 & 0.722 & 0.785 & 0.786 & 0.637 & 0.715 \\
    \#07 & \checkmark & 4 & \checkmark &- &L & 0.755 & 0.752 & 0.587 & 0.705 & 0.754 & 0.751 & 0.579 & 0.694 \\
    \hline
    \#08 & \checkmark & 4 & \checkmark &\checkmark &L$\rightarrow$L & 0.748 & 0.717 & 0.577 & 0.705 & 0.760 & 0.741 & 0.587 & 0.693 \\
    \#09 & \checkmark & 4 & \checkmark & \checkmark & L$\rightarrow$G & 0.788 & 0.780 & 0.645 & 0.741 & 0.776 & 0.768 & 0.618 & 0.715 \\
    \#10 & \checkmark & 4 & \checkmark & \checkmark & G$\rightarrow$G & 0.778 & 0.763 & 0.627 & 0.726 & 0.767 & 0.753 & 0.599 & 0.694 \\
    \hline
    \rowcolor{mygray}
    \textbf{\#OUR} & \checkmark & 4 & \checkmark & \checkmark & G$\rightarrow$L 
    & \textbf{0.806}  & \textbf{0.798}  & \textbf{0.676}  & \textbf{0.756}  & \textbf{0.797}  & \textbf{0.793}  & \textbf{0.653}  & \textbf{0.737}  \\
    \hline
    \end{tabular}
    \vspace{-5pt}
\end{table*}

\begin{figure*}[t!]
    \centering
    \includegraphics[width=\linewidth]{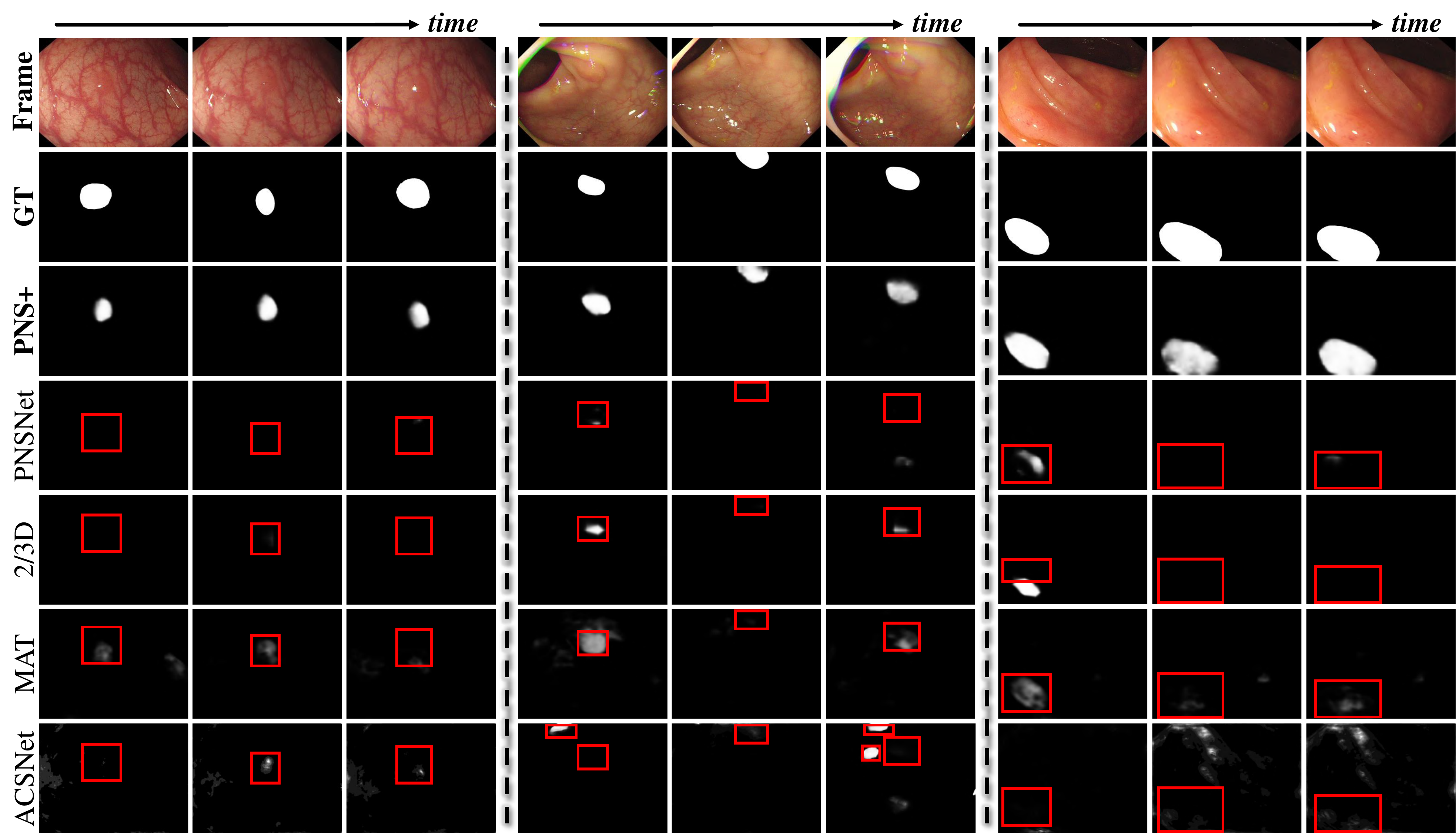}
    \caption{Qualitative visualization of the proposed \ourmodel~and four representative competitors on three sequences (from left to right: case14\_3, case30, and case3\_2).
    The red boxes indicates the wrong or missing predictions.
    We refer the readers to the project page for completed dynamic comparison.
    }\label{fig:front_figure}
\end{figure*}

\subsubsection{Contribution of Channel Split}
To discover the best setting for the channel split rule as in~\eqnref{eqn:channel_split_rule}, we instantiate four variants with four different channel split numbers: \#02 ($N$$=$$1$), \#03 ($N$$=$$2$), \#04 ($N$$=$$4$), and \#05 ($N$$=$$8$).
These results show that small (\#02 \& \#03) and large (\#05) channel split numbers may harm the channel-level information by collapsing the knowledge in a different channel.
In contrast, we adopt the moderate scale (\#04: $N$$=$$4$) with the best performance on \ourdataset-\texttt{Hard} (Unseen) (\eg, Dice: 2.7\%$\uparrow$) when compared to variant \#05.
Such a trade-off scale would exert our model focusing on the polyp-related attention while suppressing the irrelevant information.

\subsubsection{Contribution of Soft-attention}
We further ablate soft-attention and observe that \#04 with the soft-attention block is generally better than \#06 without it on \ourdataset-\texttt{Easy} (Unseen): 1.9\%$\uparrow$ in terms of Dice score.
Such improvement suggests that introducing the soft-attention operation to synthesize the relationship between aggregation feature and affinity matrix is necessary for increasing performance.

\subsubsection{Effectiveness of Normalization}
We also study the improvement of the normalization operation by comparing \#04 with \#07.
We observe that \#04 generally outperforms \#07 on \ourdataset-\texttt{Hard} (Unseen) (\eg, Dice: 4.1\%$\uparrow$).
It shows that the layer normalization along the temporal dimension could alleviate the internal covariate shift problem by fixing the distribution of query entries in the attention mechanism.

\begin{figure}[t!]
    \centering
    \includegraphics[width=\linewidth]{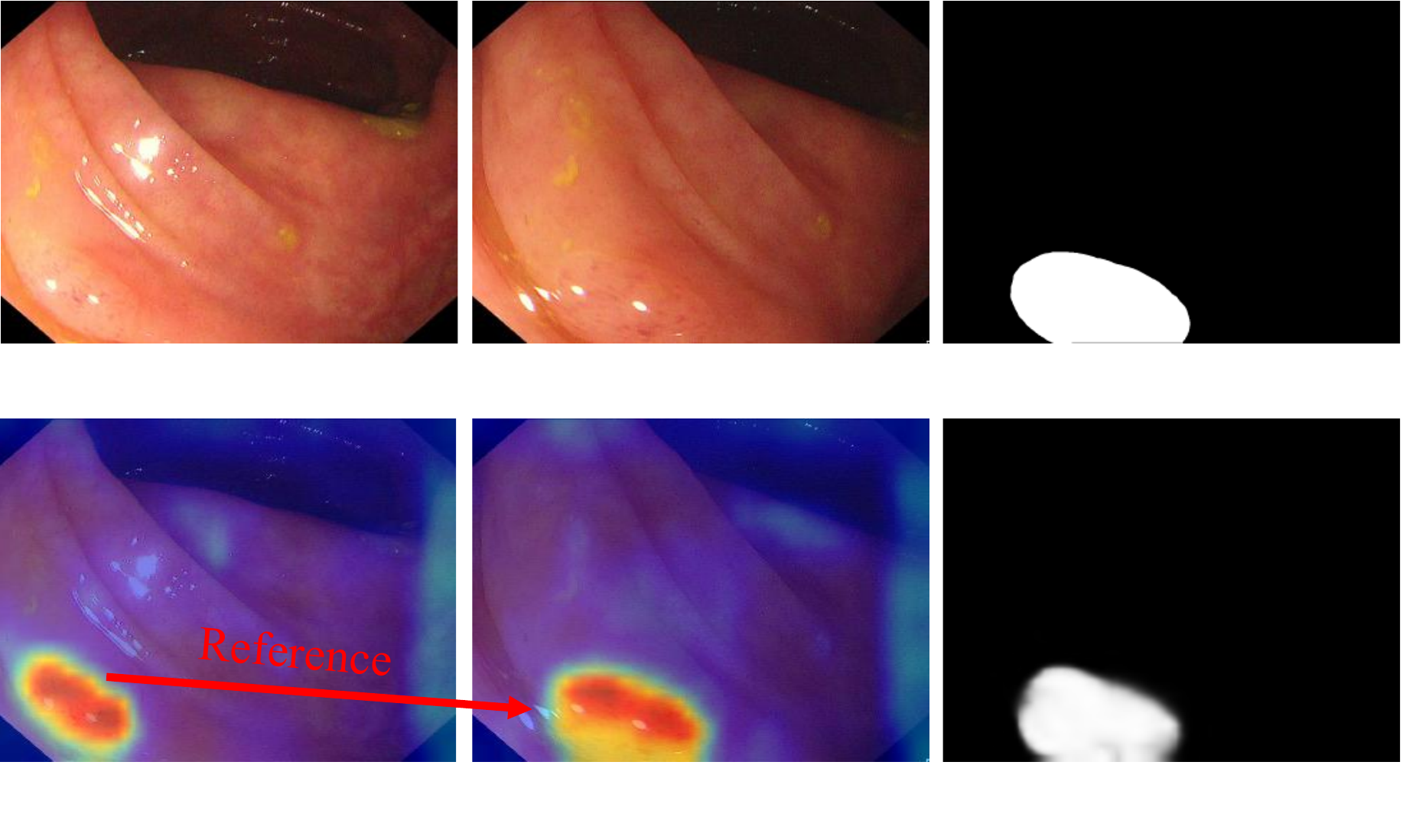}
    \put(-217, 69){\footnotesize (a) Anchor frame $I_1$}
    \put(-145, 69){\footnotesize (b) Current frame $I_s$}
    \put(-71.5, 69){\footnotesize (c) Ground truth $G_s$}
    \put(-200, 4.5){\footnotesize (d) Anchor}
    \put(-197, -5){\footnotesize feature $\mathcal{A}^h$}
    \put(-144, 4.5){\footnotesize (e) Spatial-temporal}
    \put(-120, -5){\footnotesize feature $Z^l$}
    \put(-65, 4.5){\footnotesize (f) Prediction $P_s$}
    \vspace{8pt}
    \caption{Feature visualization of key dataflows.
    The red arrow denotes using the anchor feature $\mathcal{A}^h$ to guide the representation of spatial-temporal frame $Z^l$.
    More details refer to \secref{sec:abla_learning_strategy}.}
    \label{fig:heatmap}
\end{figure}
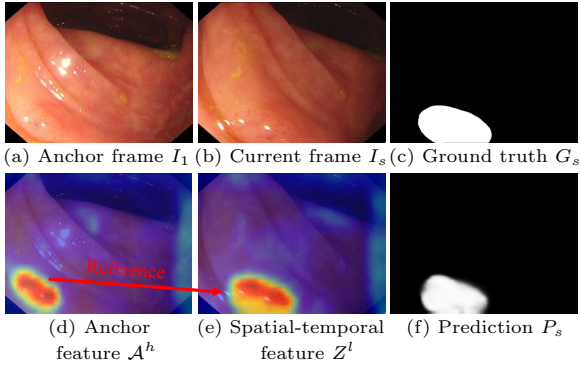

\subsubsection{Different Learning Strategies}\label{sec:abla_learning_strategy}
Finally, we examine the effectiveness of the proposed learning strategy, as proposed in \secref{sec:glns_pipeline}, by deriving three variants, including \#08 (L$\rightarrow$L: local-to-local), \#09 (L$\rightarrow$G: local-to-global), \#10 (G$\rightarrow$G: global-to-global), and \textbf{\#OUR} (G$\rightarrow$L: global-to-local).
For example, variant \#09 combines local spatial-temporal cues and introduces global ones, termed a local-to-global (L$\rightarrow$G) strategy.
\#08 will dramatically decrease on \ourdataset-\texttt{Easy} (Unseen) ($\mathcal{S}_\alpha$: 5.8\%$\downarrow$) when focusing on the local cues due to a lack of global context.
On the other hand, if only focusing on the global information, the performance of variant \#10 will drop on \ourdataset-\texttt{Hard} (Unseen), \eg, $F_\beta^{w}$:~5.4\%$\downarrow$.
In contrast, \textbf{\#OUR} with the global-to-local strategy outperforms variant \#09 on \ourdataset-\texttt{Hard} (Unseen), \eg, $F_\beta^{w}$: 3.5\%$\uparrow$, since propagating long-term cues into short-term neighbors.

We further validate the effectiveness of the global-to-local learning strategy via visualizing the key dataflows.
As shown in~\figref{fig:heatmap}, the first and second columns present the anchor feature $\mathcal{A}^h$ extracted from the global encoder and the spatial-temporal feature $Z^l$ from the second NS block, respectively.
Note that the current frame $I_s$ is randomly selected from consecutive frames $I_\Delta$.
It shows that our \ourmodel~can propagate the long-term dependency with the assistance of the anchor frame $I_1$, though the current frame $I_s$ is hard to recognize due to indefinable boundaries (\ie, IB attribute).
Of note, as in the rightmost column of \figref{fig:front_figure}, the PNSNet fails to locate the polyp since it does not use a global-to-local learning strategy.
Compared to it, our \ourmodel~successfully detects the polyp by exploiting the global reference of the anchor frame.

\begin{figure*}[t!]
    \centering
    \includegraphics[width=\linewidth]{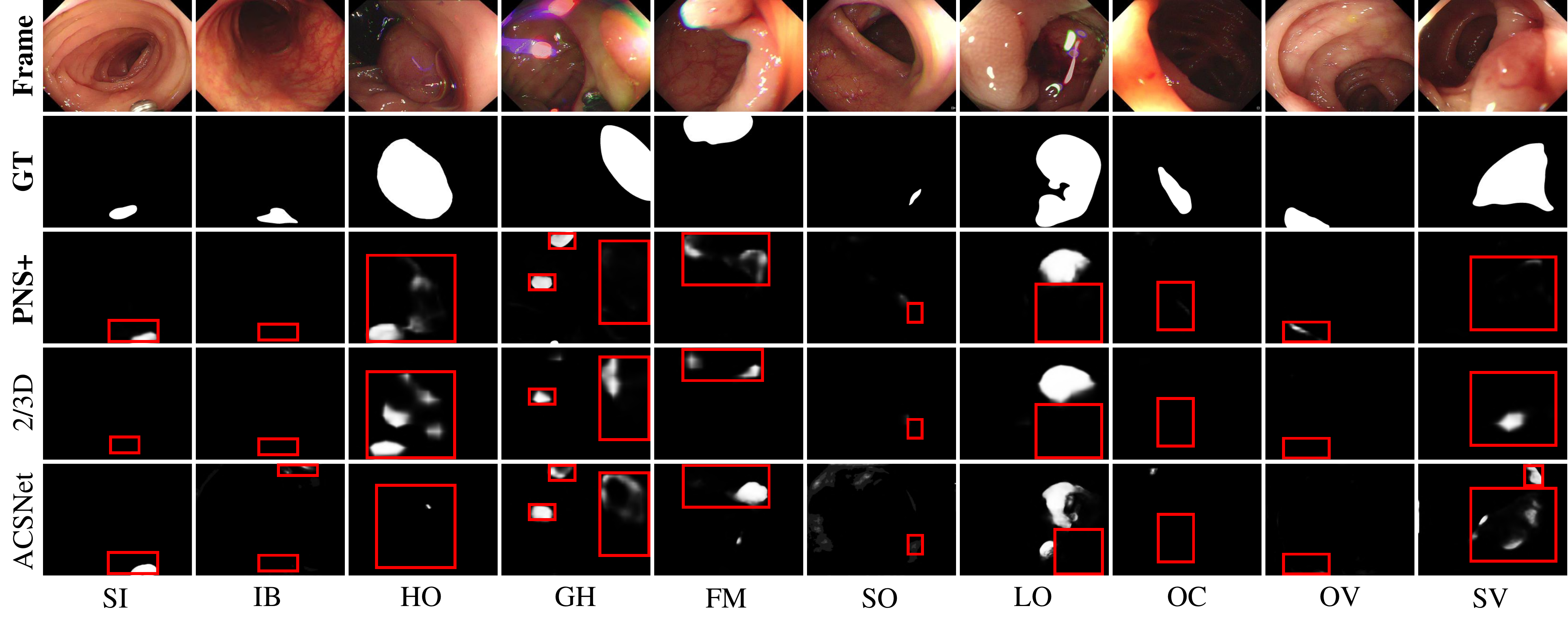}
    \caption{Challenging samples were taken from ten visual attributes.
    More analyses can be referred to~\secref{sec:failure_case}.}
    \label{fig:failure}
\end{figure*}

\subsection{Issues and Challenges}\label{sec:failure_case}
This section discusses some common issues within challenging attributes, whose visualization results are presented in~\figref{fig:failure}.
Of note, VPS is a newly-emerging and challenging track in medical imaging, and its overall accuracy is not high enough.
We observe that existing cutting-edge models (\ie, ACSNet and 2/3D) and our baseline model (\ourmodel) still lack sufficient robustness in particular cases in LO, HO, SI, GH, and SV attributes.
As for the HO ($3^{rd}$ column) and LO ($8^{th}$ column) attributes, three models fail to capture the whole polyp due to significant appearance changes.
Besides, the false-positive/-negative predictions (marked with red boxes) on the surgical instrument ($1^{st}$ column) and the optical flares ($4^{th}$ column) indicate that these models could not learn semantics without perceiving the accurate polyp-related representation in such a hard case.
%
Moreover, the misidentifications for the SV attribute (last column) are caused by the insufficient diversity of polyp sizes in the training set.
The aforementioned drawbacks inspire us to explore more robust learning paradigms to improve the accuracy of VPS.

We also observe that three models consistently fail to locate lesion regions that share a similar color with the intestinal wall or are too small to be detected.
Thus, there is a large room for improving the detection ability in IB and SO attributes via camouflaged pattern discovery techniques~\cite{fan2021concealed,ji2022gradient}.
Last but not least, lacking temporal-wise understanding will lead to the false prediction in the FM, OV, and OC attributes.
Taking OV and OC, for example, exploiting temporal cues more thoroughly should mitigate the performance degradation results from the occlusion of the intestinal wall or the image boundary since the occlusion is not continuous in the entire video clip.
To sum up, these challenging cases are the common difficulties other methods face and cause a severe performance degradation that deserves further exploration.

\section{Potential Directions}\label{sec:future_trends}
This section highlights several potential trends for promoting colonoscopy research in the deep era.
\begin{itemize}
    \item \textbf{High-precision Diagnosis.} 
    As shown in \tabref{tab:ModelScore}, we observe that the leading approaches are still unsatisfactory in our \ourdataset-\texttt{Hard} (\eg, sensitivity score $<$ 0.63).
    We argue that the high-precision VPS algorithm would steer clinical medicine in boosting auxiliary diagnostic technologies.
    
    \item \textbf{Data-insufficient Learning.} 
    It is promising to explore efficient learning strategies~\cite{guo2021semantic,senkyire2021supervised} under limited conditions in specific clinical applications, such as weakly-/un-/self-supervised learning and knowledge distillation.
    
    \item \textbf{Privacy-preserving AI.} 
    Intelligent VPS systems must safeguard data through the entire life cycle from training to production and governance, which fuels fundamental techniques like federal learning.
    
    \item \textbf{Trustworthy AI.} 
    How AI-guided decisions are made and what determining factors are involved play a crucial role in understanding the insights of deep networks.
    In other words, the VPS model should be causal, transparent, explainable, and interactive, which inspires more trusted developments, such as~\cite{zou2022tbrats}.
\end{itemize}

The above possible directions listed are still far from being solved for the VPS.
Fortunately, several famous works can be served as references, providing it a potential basis to be transferred to our community.

\section{Conclusion}
This paper presents the first comprehensive study on video polyp segmentation (VPS) from a deep learning perspective.
We first introduce a large-scale VPS dataset \ourdataset~via extending the famous SUN-database with diversified annotations, \ie, attribute, object mask, boundary, scribble, and polygon.
We then design a simple but efficient baseline, dubbed \ourmodel, to segment colon polyps from the colonoscopy video.
Based on the normalized self-attention block, \ourmodel~fully exploits long-term and short-term spatial-temporal cues via a novel global-to-local learning strategy.
We further contribute the first comprehensive benchmark containing 13 cutting-edge polyp/object segmentation approaches.
Extensive results show that \ourmodel~achieves the best performance against all these competitors.
We conclude by outlining several potential directions for future colonoscopy-related research in the deep learning era.
We hope this work will spur advancements in other closely related medical video analyses.

\section*{Acknowledgments}
The authors would like to thank the anonymous reviewers and editor for their helpful comments on this manuscript.
Besides, we thank Huazhu Fu for his insightful feedback.

\section*{Conflicts of Interests}
The authors declared that they have no conflicts of interest in this work.
We declare that we do not have any commercial or associative interest that represents a conflict of interest in connection with the work submitted.

\bibliographystyle{IEEEtran}
\bibliography{mir-article}

\end{document}